\documentclass[aps,nofootinbib,notitlepage,superscriptaddress,twocolumn,10pt,prd]{revtex4-1}

\usepackage{comment}
\usepackage{graphicx}
\usepackage{amsmath,amssymb}
\usepackage{hyperref}
\usepackage{braket}
\usepackage{subfigure}
\usepackage{float}
\usepackage[dvipsnames]{xcolor}
\usepackage{soul}
\usepackage{rotating}
\usepackage{multirow}
\usepackage{mathtools}
\usepackage{makecell}

\usepackage[margin=0.75in]{geometry}

\hypersetup{
	colorlinks=true,
	linkcolor=red,
	citecolor=blue,
}

\usepackage{amsthm}

\theoremstyle{definition}

\usepackage{xcolor}
\definecolor{orange}{rgb}{1,0.5,0}

\newcommand{\rref}{\mathrm{ref}}
\newcommand{\ra}{\mathrm{ra}}
\newcommand{\dec}{\mathrm{dec}}
\newcommand{\Ngs}{N_\mathrm{gs}}
\newcommand{\KL}{\mathrm{KL}}
\newcommand{\ML}{\mathrm{ML}}
\newcommand{\JN}{{J\mspace{-2mu}  N}}
\newcommand{\dH}{\mathrm{H}}
\newcommand{\dL}{\mathrm{L}}
\newcommand{\dV}{\mathrm{V}}

\usepackage{natbib}

\begin{document}

\title{Machine Learning Assisted Parameter-Space Searches for Lensed Gravitational Waves}

\author{Giulia Campailla}
\affiliation{Department of Physics and INFN, University of Genova, Via Dodecaneso 33, 16146, Italy}

\author{Marco Raveri}
\affiliation{Department of Physics and INFN, University of Genova, Via Dodecaneso 33, 16146, Italy}

\author{Wayne Hu}
\affiliation{Kavli Institute for Cosmological Physics, Enrico Fermi Institute, and Department of Astronomy \& Astrophysics, University of Chicago, Chicago IL 60637, USA}

\author{Jose Mar\'ia Ezquiaga}
\affiliation{Center of Gravity, Niels Bohr Institute, Blegdamsvej 17, 2100 Copenhagen, Denmark}

\begin{abstract}	
When a gravitational wave encounters a massive object along the line of sight, repeated copies of the original signal may be produced due to gravitational lensing.
In this paper, we develop a series of new machine-learning based statistical methods to identify promising strong lensing candidates in gravitational wave catalogs.
We employ state-of-the-art normalizing flow generative models to perform statistical calculations on the posterior distributions of gravitational wave events that would otherwise be computationally unfeasible. Our lensing identification strategy, developed on two simulated gravitational wave catalogs that test noise realization and event signal variations, selects event pairs with low parameter differences in the optimal detector basis that also have a high information content and favorable likelihood for coincident parameters.
We then apply our method to the GWTC-3 catalog and find a single pair still consistent with the lensing hypothesis. This pair has been previously identified through more costly evidence ratio techniques, but rejected on astrophysical grounds, which further validates our technique.
\end{abstract}
	
\maketitle

\section{Introduction}\label{sec:introduction}

As they travel through the Universe, gravitational waves (GWs) are only affected by the gravitational interactions with the cosmic structures.
In the limit of weak-gravity and geometric optics, gravitational lensing of GWs obeys the same principles as optical lensing~\cite{Schneider:1992}. 
This implies that signals will be deflected and delayed and, in cases of good alignment between the source, lens and observer, even produce multiple images. 
In the context of GWs produced by the coalescence of compact binaries, repeated chirps from a single merger will arrive at the detectors.
Because GWs are coherently emitted and detected, gravitational lensing of GWs has the advantage of retaining phase information in the waveform.

In the strong lensing regime, multiple copies of the same gravitational wave event $h(t)$ are produced. 
Depending on the path through the lens, each copy $j$ experiences separate modifications to its amplitude (through the magnification factor $\mu_j$), arrival time at the detector ($t_d^j$), and phase~\cite{Blandford:1986zz}. 
These lensed signals can be expressed in Fourier space in terms of the original signal $\tilde h(f;t_d)$ as:
\begin{align} \label{Eq:GW_strong_lensing}
&\tilde{h}_\mathrm{lensed}^{j}(f;t_d^j) = \sqrt{|\mu_j|} e^{i n_j \pi / 2} \tilde{h}(f;t_d^j) \,, \nonumber \\
&t_d^j = t_d + \Delta t_j \,,
\end{align}
where the Morse phase shift of $n_j \pi / 2$ depends on the type of solution of the lens equation with $n_j = 0, 1, 2$ corresponding to type I, II, and III images.
Here $\Delta t_j$ represents the time delay relative to the unlensed signal.
In this regime gravitational lensing does not modify the frequency evolution of the signal, but the overall Morse phase may induce small waveform distortion in the presence of multiple harmonics with different frequency evolution~\cite{Dai:2017huk,Ezquiaga:2020gdt}.
Frequency-dependent modifications may be possible when the source is close to a caustic in the high-magnification limit~\cite{Lo:2024wqm,Ezquiaga:2025gkd}, or when it encounters some substructures in the wave optics regime~\cite{Nakamura:1999uwi,Takahashi:2003ix}. 
However, since they require a different search strategy, we do not consider those here.

For lensed GWs, we cannot spatially resolve the images because of the limited localization accuracy of current detector networks.
Typical image separations for a galaxy-scale lens are at the level of arcseconds, while the best current GW sky localizations only go down to a few square degrees.  
On the other hand, ground-based GW detectors can temporally resolve the images, with a resolution of milliseconds. 
This is orders of magnitude more precise than for most electromagnetic transients, and enables searches across a very wide range of lenses, from compact stellar mass lenses to the largest galaxy clusters~\cite{Vujeva:2025kko}. 

Here we focus on searches for lensed GWs in the regime of multiple images/repeated chirps, where the time delay is much longer than the duration of the strain signal, so that the signals do not overlap temporally.
For binary black holes detected by the LIGO-Virgo-KAGRA detectors~\cite{LIGOScientific:2014pky,VIRGO:2014yos,KAGRA:2020tym}, this effectively includes any lens producing time delays longer than $\sim\! 10$\,s.

The challenge, however, is to determine whether separate chirps in the strain are independent, unrelated events or different images of the same strongly lensed GW signal~\cite{Caliskan:2022wbh}.
Identification of lensing candidates requires a catalog search for all pairs that could possibly be lensed copies of each other.
Current searches have not yet found evidence for strongly lensed GWs ~\cite{Hannuksela:2019kle,LIGOScientific:2021izm,Janquart:2023mvf,LIGOScientific:2023bwz}, but such events are expected to appear in GW catalogs once they reach thousands of events~\cite{Ng:2017yiu,Li:2018prc,Xu:2021bfn}.

Given Eq.~\eqref{Eq:GW_strong_lensing}, one could conduct the search for pairs directly using the strain data.
In fact, several machine learning methods have been proposed as low-latency tools to analyze the time-frequency spectrograms~\cite{Goyal:2021hxv,Magare:2024wje}.
The challenge here is to efficiently train the networks in actual non-stationary and non-Gaussian noise.
The next level would be looking at the waveform morphology, for example using cross-correlations of the maximum likelihood waveform at each detector~\cite{Chakraborty:2024mbr}.
However, this is complicated by the different noise realizations and changes in signal and noise properties due to the difference in orientation and arrival times at the detectors.
This approach is also statistically inefficient, as the two noisy waveforms will differ by random fluctuations in many noise degrees of freedom while being characterized by only a handful of physical properties that have to match, increasing the chance of false matches from coincidental noise.

On the other hand, given their origin in a common merger event, waveform model parameters must match, compatibly with Eq.~\eqref{Eq:GW_strong_lensing}.
In this case model parameters can be thought as an efficient form of data compression to the extent that waveform approximants provide a good description of the waveform.
Typically the number of parameters is much smaller than the number of data points and correspond to identifiable physical properties of the event.  Looking for lensing consistency in terms of these parameters therefore minimize the look-elsewhere effect of a posteriori checks of pair consistency.

The most direct strategy would be to compute the overlap of the posterior samples of pairs of events.
Computationally more costly, one could also perform a joint parameter estimation of a pair of events to compute the likelihood of lensing through the evidence ratio of the lensed and not-lensed hypotheses\ \cite{Haris:2018vmn,Barsode:2024zwv,Lo:2021nae,Janquart:2021qov}. 
The information on interesting lensing candidates can be used to perform targeted searches of sub-threshold events to find additional copies of the same source~\cite{Li:2019osa,Li:2023zdl,Goyal:2023lqf}, or to cross-match with existing catalogs of strong lenses to identify electromagnetically potential lenses or host galaxies~\cite{Vujeva:2024scq}. 
Alternatively, a strongly lensed event could be identified as a population outlier if the magnification is large enough to significantly change the inferred masses, although this is rare given current measurement uncertainties, noise fluctuations and the imperfect knowledge of the intrinsic astrophysical population~\cite{Farah:2025ews}.

Even parameter consistency checks entail significant challenges.
State-of-the-art waveform approximants involve a large number of partially degenerate parameters, and performing consistency checks with high-dimensional, multimodal, non-Gaussian distributions is so technically demanding that it threatens practical feasibility.
To perform statistical calculations with these distributions, Kernel Density Estimators (KDE) are often used but are known to suffer from the curse of dimensionality~\cite{10.1214.aos.1176345206,091533b0-fe38-3eea-b846-fd10afc72174,Raveri:2021wfz}, requiring a number of samples that grows exponentially with the number of parameters to achieve a given accuracy.
This problem can be alleviated by choosing a smaller set of parameters that are most closely related to the observable properties of the events which are relevant for lensing~\cite{Ezquiaga:2023xfe} but can never be completely avoided.
There are also other, less fundamental, technical challenges: the parameter space  contains many parameters that are periodic; the prior distribution is not uniform on many parameters.

In this paper we present several methodological advancements to overcome these challenges.
Instead of KDE estimators we use normalizing flow models specifically tailored to accurately approximate the posterior distributions of GW event parameters.
We then develop the statistical methods to perform lensing consistency tests in parameter space, which include the calculation of the information content of a pair of events, the quantification of the significance of a parameter difference, fully taking into account non-Gaussianities of posterior distributions, and the likelihood ratio for the lensing hypothesis.
We validate our methods on simulated catalogs of lensed and unlensed events, and then apply them to the public LVK catalog, GWTC-3.

This paper is organized as follows:
in Sec.~\ref{sec:lensing} we discuss the parameter space and the parameter bases employed to perform lensing consistency tests;
in Sec.~\ref{sec:nf.modelling} we describe the normalizing flow models used to approximate the posterior distributions of GW events (readers not interested in the technical details of the machine learning architecture may safely skip this section);
in Sec.~\ref{sec:statistical.methods} we introduce the statistical methods applied throughout this work;
in Sec.~\ref{sec:datasets} we outline the datasets that we use to validate our methods, including two simulated catalogs and the GWTC-3 catalog;
in Sec.~\ref{sec:results.info} we show how to choose a parameter basis for lensing searches based on the information content that it captures (readers primarily interested in strong-lensing consistency results may proceed directly to Secs.~\ref{sec:results.identification}--\ref{sec:results.real});
in Sec.~\ref{sec:results.identification} we present lensing identification results on simulated catalogs, illustrating the development, testing, and validation of our approach , and propose an identification strategy for lensing candidates based on these results;
in Sec.~\ref{sec:results.real} we apply our methods to the GWTC-3 catalog and report the results.
We conclude in Sec.~\ref{sec:conclusions}.  
Appendix \ref{App:Math.Details} provides mathematical details for the statistical properties of our methods and Appendix \ref{app:real.lensing.metrics} details the agreement between methods for events in GWTC-3. 

\section{Lensing tests in parameter space}\label{sec:lensing}

The core of our analysis relies on the parameter posterior distribution for each GW event.
We denote this with $P(\theta | d, M)$, where $\theta$ represents the parameter vector, $d$ stands for the observed strain data and $M$ the waveform and noise model.
We only consider GW events from binary black hole mergers (BBH) and, since the model $M$ is fixed, we drop $M$ in the following so as not to be confused with mass. 

In gravitational-wave parameter estimation~\cite{Christensen:2022bxb}, the posterior probability density is computed using Bayes' theorem:
\begin{align} \label{Eq:Bayes_probability_densities}
P(\theta | d) = \frac{P(d | \theta) P(\theta)}{P(d)} \equiv \frac{\mathcal{L}(d | \theta) \Pi(\theta)}{\mathcal{E}(d)},
\end{align}
where $\mathcal{L}(d | \theta)$ is the likelihood, representing the probability of the detectors measuring the data $d$, given an event with source properties $\theta$ that parameterize the  waveform approximant; $\Pi(\theta)$ is the prior, encoding any a priori knowledge or assumptions about the parameters $\theta$ under the model considered; $\mathcal{E}(d)$ is the evidence, which normalizes the posterior, and it is defined as:
\begin{align} \label{Eq:evidence}
   \mathcal{E}(d) = \int \mathcal{L}(d | \theta) \Pi(\theta) \, d\theta \,.
\end{align}
The standard parameter estimation framework for BBH events typically involves 15 parameters for non-eccentric binaries.
These are divided into 8 intrinsic parameters, $\{m_1,m_2,\vec{S}_1,\vec{S}_2\}$ for the masses, $(m_1, m_2)$, and spins of the objects in the binary, $(\vec{S}_1, \vec{S}_2)$, and 7 extrinsic parameters, $\{\ra,\dec,\psi,\iota,d_L,t_\rref,\phi_\rref\}$ which describe position on the sky using the right ascension and declination $(\ra,\dec)$, polarization orientation $(\psi)$, inclination $(\iota)$, luminosity distance $(d_L)$, arrival time $(t_\rref)$ and reference phase $(\phi_\rref)$ of the signal~\cite{Christensen:2022bxb}.
The intrinsic parameters can alternately be represented by  the detector frame chirp mass $\mathcal{M}_c=(m_1m_2)^{3/5}/(m_1+m_2)^{1/5}$, the mass ratio $q$ between the primary and secondary masses, and the two spin components, consisting of two spin magnitudes $a_1,a_2$ and their 4 relative angles. 
Likewise, the orbital inclination can be instead represented by the angle between the total angular momentum and the line of sight $\theta_\JN$.  
We adopt standard priors for these parameters, as implemented for LVK analyses~\cite{stronglensingligovirgo2023}. 

When comparing the parameter posteriors for two events, we marginalize the posterior distributions over the luminosity distance and the arrival time as we expect them to differ because of magnification and time delay between two lensed images.
This marginalization reduces the parameter space to 13 dimensions.

We study lensing consistency in two alternative parameter bases to reduce dimensionality further, effectively compressing the data:
\begin{itemize}
\item {\bf Detector basis}: this basis was first introduced in~\cite{Ezquiaga:2023xfe} and contains 6 parameters. 
It includes a phase parameter for each detector, namely Hanford, Livingston, and Virgo denoted $\phi_\dH,\phi_\dL,\phi_\dV$ respectively (`detector phases').
These encode the phase of the 22-mode at a fiducial frequency, with default value of 40\,Hz, at each detector. Each of these parameters is periodic with range $[0, 2\pi]$.
Two parameters, $\tau_{\mathrm{HL}}$ and $\tau_{\mathrm{LV}}$, quantify the time delay between the arrival times at each of the detectors, which encode $\ra,\dec$ information, whereas $\Delta\phi_f$ gives the number of cycles between  20-100\,Hz and, mostly, takes the place of $\mathcal{M}_c$.  When comparing two events, we must pick the detector basis of one event and transform the localization of the second event onto the common basis, as if the detectors had made their time delay measurements from the positions of the first event
(see \cite{Ezquiaga:2023xfe} for details).
\item {\bf Overlap basis}: this parameter basis consists of 9 parameters, commonly used in LVK posterior overlap analyses~\cite{Hannuksela_2019}. These parameters are specifically  $m_{1,2}$, $\ra$, $\sin(\dec)$, $a_{1,2}$, the cosines of the polar angles of the spins with respect to the orbital angular momentum $\cos (\theta_{a1,a2})$, and $\theta_\JN$. 
\end{itemize}
In the overlap basis, following~\cite{Hannuksela_2019}, we do not include phase parameters such as the binary coalescence phase at a reference frequency, $\phi_{\rm ref}$, and the polarization angle, $\psi$ which impacts the detected phase.  
This effectively marginalizes over any possible Morse phase shift introduced by lensing. The reason for this choice is that beyond the 22-mode of a circular binary, a shift in $\phi_{\rm ref}$ does not correspond exactly to a Morse phase shift~\cite{Barsode:2024zwv}. For this reason we do not include the detector phases in the overlap basis, compatibly with~\cite{Hannuksela_2019}.

The detector basis parameters evades some of these issues by first reconstructing the $22$ mode from the original 15-dimensional space before transforming it to the detector basis.   
For a given event, the 3 detector phases encode 
$\{\phi_{\rm ref},\psi,\theta_\JN\}$ in a manner that reduces  degeneracies between the parameters and with other poorly determined parameters such as spin components and mass ratio.
It also naturally includes the possibility of a Morse phase  that shifts the three detector phases all by the same quantity.  
When considering any statistic for the lensing consistency between two events in this basis, we maximize this probability over the 4 Morse phase difference possibilities $0, \,\pm \pi/2,\, \pi$.

On the other hand, when comparing two events we can only match the detector basis of one of the events since in general the Earth has rotated the detector positions given the time delay.  
Since a lensed event should be consistent in either detector basis projection, we take the smaller of the two consistency probabilities for each test.

For all bases, we propagate the priors consistently from the original 15-dimensional space.
This is a significant difference between this analysis and the standard LVK implementation that assumes uniform priors with fixed ranges for parameters in the overlap basis~\cite{Garr_n_2023,Haris:2018vmn,Barsode:2024zwv}.
We note that, in the standard LVK analysis, the prior distribution is typically different and adapted to each event, in particular the masses, which we propagate consistently for each event in the pair.

\section{Normalizing Flow models for GW posteriors}\label{sec:nf.modelling}

We address the challenge of performing computationally intensive statistical calculations required for assessing lensing parameter consistency between GW events by utilizing normalizing flows (NFs). 
These machine learning models~\cite{Papamakarios:2016ctj, Kingma:2016wtg, Rezende:2015ocs, Papamakarios:2017tec} transform simple base distributions, such as a Gaussian, into complex target distributions through a series of invertible and differentiable mappings, enabling efficient density estimation and sampling.
When the posterior is either unavailable or slow to evaluate, NFs provide a fast and accurate approximation of the posterior distribution. When the dimensionality of the problem exceeds a few parameters,  NFs outperform kernel density estimation methods in both accuracy and computational efficiency, as shown in~\cite{Raveri:2021wfz}.

We denote with $\theta^\prime$ the coordinates of an abstract latent space, where we consider a base distribution $P^\prime(\theta^\prime)$. This base distribution is typically chosen to be analytically tractable; here, we take it to be a multivariate Gaussian distribution centered at zero with unit variance and no off-diagonal covariance. 
If we denote with $\theta$ the physical parameter space where the  target distribution is defined, we can connect the two spaces with the mapping $\theta = f(\theta^\prime)$.
If this mapping is bijective it would transform a distribution in the latent space, defined over $\theta^\prime$, into a distribution defined over the physical space $\theta$ as:
\begin{align} \label{Eq:CoordinateTransformationProbability}
    \log P(\theta) = \log P^\prime(\theta^\prime) + \log \left| \det \frac{\partial f}{\partial \theta^{\prime}} \right| \,,
\end{align}
where log denotes the natural logarithm, $|\cdot|$  the absolute value and $\det (\partial f / \partial \theta^\prime)$ the determinant of the Jacobian of the transformation $f$. 

A working NF builds $f$ in such a way that the transformed distribution matches the target distribution.
This is achieved in practice by defining the transformation as a composition of several invertible transformations, each of which will depend on many trainable parameters.
These parameters are optimized to minimize the difference between the target input distribution and the transformed distribution.

The inverse transformation $f^{-1}$ enables the mapping of the target distribution back into the latent space, effectively transforming it into a multivariate Gaussian distribution. 
This inverse operation lends a double meaning to its name despite its general applicability to any analytically tractable base distribution defined in the latent space.

It has been shown in~\cite{NIPS2000_3c947bc2} that such a mapping exists for any target distribution.
Once the mapping $f$ and its Jacobian are known, the probability density can be evaluated using Eq.\ (\ref{Eq:CoordinateTransformationProbability}).
Furthermore, sampling from the target distribution is achieved by first sampling from the latent distribution and then transforming these samples into the physical parameter space via $f$.

We use the implementation of NFs in {\tt tensio\-meter} \cite{Raveri:2018wln,Raveri:2021wfz}, which we have extended to accommodate the specific requirements of gravitational wave posterior distributions.
A key feature of these flow models is their capability to handle periodic boundaries, a necessity for certain parameters, such as angular coordinates, commonly encountered in GW posteriors. 
This is achieved through the use of spline-based flows with periodic boundary conditions, as outlined in~\cite{2019arXiv190604032D, 2020arXiv200202428J}.

The flow architecture we employ is constructed as follows. 
We begin by applying a series of fixed pre-processing transformations designed to ensure its suitability for subsequent modeling. 
These pre-processing steps include scaling to standardize parameters, linear decorrelation of non-periodic parameters to minimize covariance and improve statistical independence, and wrapping of periodic parameters to re-center them within their periodic domains.
The trainable part of the normalizing flow stacks $2 \log_2 N_{\rm params}$ spline flows, with fixed permutations between layers.

One of the unique features of NFs implemented in the {\tt tensiometer} framework is multiscale learning rate adaption, described in~\cite{Raveri:2024dph}.
This allows the flow to learn increasingly detailed features of the distribution and automatically conclude training once they have all been learned, ensuring both computational efficiency and robustness.
Following~\cite{Raveri:2024dph}, we train several flows and average them to safeguard against catastrophic weight initialization and improve flow performances in regions with sparse samples.
\begin{figure}[!ht]
\centering
\includegraphics[width=\columnwidth]{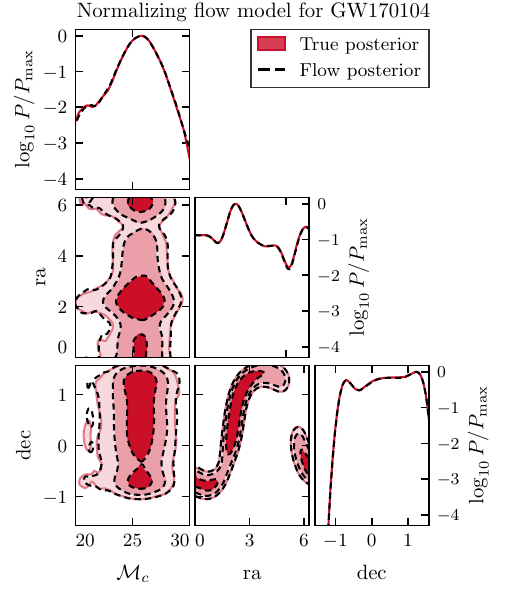}
\caption{ \label{fig:flow.poster}
Joint marginalized posterior distribution for selected parameters of the GW170104 event.
The full posterior, and flow model, contain 15 parameters which are sampled to produce the 2D and 1D distributions shown here.
In joint 2D panels, the contours represent the 68\%, 95\% and 99.7\% credible regions of the posterior distribution.
The diagonal panels show the 1D marginalized posterior distributions for the same parameters. 
Log scale is used to highlight the agreement of the distribution and flow model on the tails of the distribution.
}
\end{figure}
In Fig.~\ref{fig:flow.poster} we showcase the typical performances of a flow model by showing the joint 2D and 1D marginalized posterior distributions for one GW event. 
The NF model is trained on the full 15-dimensional posterior distribution of the GW170104 event and only three selected parameters are shown.
As we can see the sampled distribution and the flow model match very closely, even for extremely non-Gaussian parameters.
The 1D distributions in Fig.~\ref{fig:flow.poster} are shown  on a log scale to highlight that the flow model is able to capture accurately the tails of the distribution.

To further assess the quality of the learned distribution we perform a Kolmogorov-Smirnov (KS) test in the latent space using validation samples. This statistical test quantifies the degree to which the validation samples conform to the expected distribution, specifically a multivariate Gaussian in latent space. By comparing the empirical cumulative distribution function (CDF) of the samples, to the theoretical CDF of the Gaussian, the KS test provides a rigorous, quantitative measure of the likelihood that the validation samples originate from the target distribution.

Notably, this test is rarely performed in practical applications due to the stringent criteria required for passing~\cite{Coccaro:2023vtb}. 
Achieving a favorable outcome in the KS test demands an exceptionally close match between the learned and target distributions, particularly in high-dimensional spaces and with large sample sizes, where even minor deviations can be amplified. 

The inclusion of such a test offers a robust, additional layer of validation for assessing the fidelity of our normalizing flow model. When considering all flows that we train we find that $ 90\%$ have a KS test outcome better than $5\%$.
This result is very close to the ideal case of $95\%$ flow models having a KS test outcome better than $5\%$.

\section{Statistical Methods}\label{sec:statistical.methods}

In this section we discuss the main statistical methods of our analysis.
In Sec.~\ref{sec:information.content} we introduce how to quantify the information content of posterior distributions and how to compare them.
In Sec.~\ref{sec:parameter.shift} we discuss how to quantify the consistency of parameter determinations from two posterior distributions, as a measure of lensing consistency.
In Sec.~\ref{sec:likelihood.ratio} we show how the ratio of likelihoods at maximum posterior can be used to quantify the consistency of two posterior distributions.
Finally, in Sec.~\ref{sec:statistical.significance} we discuss how we report the statistical significance of lensing consistency tests.

\subsection{Quantifying information content}\label{sec:information.content}

To compare different parameter bases and assess their ability to capture the information present in the posteriors, we compute the Kullback-Leibler (KL) divergence between the prior and posterior distributions. 
The KL divergence, $D_{\rm KL}$, is defined as:
\begin{align} \label{Eq:Def_KL_div}
    D_{\text{KL}}(P || \Pi) = \int P(\theta) \log \left(\frac{P(\theta)}{\Pi(\theta)}\right) \, \text{d}\theta,
\end{align}
where $P(\theta)$ and $\Pi(\theta)$ are the posterior and prior distributions respectively.

Informative data impose constraints on the model parameters, leading to a difference between the prior and posterior distributions. 
The KL divergence provides a natural statistic for quantifying this change, as it measures the degree of difference between two probability distributions. Specifically, it computes the expected amount of additional information needed to describe one distribution when assuming the other. 
By choosing the prior as the reference distribution, the KL divergence effectively quantifies the informational gain contributed by the data. 

Several key properties illustrate why the KL divergence is a robust measure.
First, the KL divergence is always non-negative and equal to zero only when the two distributions coincide.
However, the KL divergence is not a metric for probability distributions because it is not symmetric.
In our context this is not a problematic aspect since there is a clear hierarchy between the prior and the posterior: the posterior updates prior knowledge with new data information and not vice versa.

Next, the KL divergence between two probability distributions cannot increase under any transformation of the parameter space $(\theta \to \theta')$. 
This property of the KL divergence is known in the context of information theory literature as the ``Data Processing Inequality''~\cite{2011arXiv1107.0740B}:
\begin{align} \label{Eq:DPI}
D_{\text{KL}}(P(\theta^\prime) \,||\, \Pi(\theta^\prime)) \leq D_{\text{KL}}(P(\theta) \,||\, \Pi(\theta)).
\end{align}
The inequality implies that any transformation or re-processing of data cannot make two distributions more distinguishable than they were in the original parameter space. In essence, it reflects the fundamental constraint that information cannot be increased through the manipulation of data.
We discuss a schematic proof, adapted from information theory literature, in App.~\ref{App:KL.Proofs}.

In the special case where the parameter transformation is one-to-one, the KL divergence remains invariant, saturating the inequality in Eq.~\eqref{Eq:DPI}.
This invariance property implies that the KL divergence is a coordinate-free measure of the difference between two probability distributions.

As a concrete example, if we assume that the posterior and prior distributions are multivariate Gaussian and we denote with $\theta_p$ and $\theta_\Pi$ their means and $\mathcal{C}_p$ and $\mathcal{C}_\Pi$ their covariances, then the KL divergence reads:
\begin{align} \label{Eq:KLGaussians}
D_{\rm KL}(P || \Pi) = &\frac{1}{2}\bigg[ (\theta_{\Pi} - \theta_p)^T \mathcal{C}_{\Pi}^{-1} (\theta_{\Pi} - \theta_p)+\\ \notag
&+\log |{\det}[ \mathcal{C}_p^{-1} \mathcal{C}_{\Pi} ]| - N_{\rm eff}\bigg],
\end{align}
where we defined
\begin{align} \label{Eq:Neff}
N_{\rm eff} &= N -{\rm tr}[ \mathcal{C}_\Pi^{-1}\mathcal{C}_p ] \,,
\end{align}
where $N$ is the total number of parameters. $N_{\rm eff}$ can be thought of as the effective number parameters that are  constrained by data.
In the Gaussian Linear Model, discussed in~\cite{Raveri:2018wln}, $N_{\rm eff}$ controls the number of degrees of freedom of the goodness of fit of the maximum posterior.

In this case, we can directly relate the KL divergence to another measure of information commonly used in the literature. This is the parameter space volume that is enclosed in a given probability iso-contour.
Assuming a Gaussian posterior, this volume is given by:
\begin{align} \label{Eq:volume}
V_p &\propto  \sqrt{\det[\mathcal{C}_p]} \,,
\end{align}  
where the proportionality constant depends on dimensionality and the iso-probability level that is chosen but drops out of volume ratios.
This measure of constraining power is used for GW posteriors in~\cite{Ezquiaga:2023xfe}, and has been commonly used in cosmology literature as, for example, a Figure of Merit for Dark Energy constraints~\cite{Albrecht:2006um}.

To relate the volume statistic to the KL divergence, we first rewrite Eq.~(\ref{Eq:KLGaussians}) as:
\begin{align} \label{Eq:KLGaussians_volume}
2D_{\rm KL}(P || \Pi) ={}& (\theta_{\Pi} - \theta_p)^T \mathcal{C}_{\Pi}^{-1} (\theta_{\Pi} - \theta_p) \nonumber \\ 
&+ 2 \log ( V_p^{-1}V_{\Pi})  - N_{\rm eff},
\end{align}
and note that we have two constant contributions in the second line and a term, in the first line, that depends on a specific data realization, $x$.
We can then average the KL divergence over data realizations and in App.~\ref{App:KL.Proofs} we discuss how to compute this expectation in the context of the Gaussian Linear model.
This gives:
\begin{align} \label{Eq:avg.KL.volume}
\langle D_{\rm KL}(P || \Pi)\rangle_x = \log V_{\Pi} - \log V_p 
\end{align}
and shows that volume measurement and the KL divergence are closely related as a measurement of constraining power.
Note that the posterior always occupies less volume than the prior so that Eq.~(\ref{Eq:avg.KL.volume}) is positive.
Also note that
the right hand side is invariant under a one-to-one parameter transformation under the Gaussian Linear Model where the Jacobian is taken to be constant, reflecting the more general parameter invariance of the  KL divergence itself.

In general it is not  feasible to calculate the data expectation of the KL divergence, but  this result establishes its close relation to parameter posterior volume.   We shall see in Sec.~\ref{sec:infoandvolume} that when we average over noise-varying data realizations of the same high signal to noise event in simulations, where the Gaussian Linear Model is a good approximation, that this expectation also holds to good accuracy.
This motivates our use of the KL divergence as the non-Gaussian extension of the figure of merit for informative data.

With a normalizing flow model for both posterior and prior distributions we can efficiently compute the KL divergence in Eq.~(\ref{Eq:Def_KL_div}) as a Monte Carlo integral:
\begin{align} \label{Eq:MC_KL_div}
D_{\rm KL}(P || \Pi) \simeq \frac{1}{N_{\theta_i}}\sum_{\theta_i \sim P} \log P(\theta_i) - \log \Pi(\theta_i).
\end{align}
Note that it is crucial to be able to both sample and calculate probability values from the posterior and prior distributions, as the normalizing flows provide.

\subsection{Parameter shifts} \label{sec:parameter.shift}

To quantify the consistency of the parameter determinations from two posterior distributions, we discuss here how to calculate the probability to exceed the likelihood of zero parameter difference.

As discussed in~\cite{Raveri:2021wfz}, we start by building the posterior distribution of parameter differences.
Following the standard approach to GW parameter estimation~\cite{Christensen:2022bxb}, we treat each event, and in particular two given events, denoted as $1$ and $2$, as independent.
Under the assumption that the two parameter sets describing the two events are different, the joint distribution of their parameter determinations is given by the product of the two parameter posteriors, since the events are independent:
\begin{align} \label{Eq:join.parameter.posterior}
P(\theta_1,\theta_2|d_1, d_2)=P_1(\theta_1|d_1)P_2(\theta_2|d_2).
\end{align}
To obtain the distribution of parameter differences we can change variables and define $\Delta\theta \equiv \theta_1 -\theta_2$ whose distribution is given by marginalizing over one of the parameters:
\begin{align} \label{Eq:param.diff.pdf}
P(\Delta \theta) = \int P_{1}(\theta) P_{2}(\theta-\Delta \theta) \, \text{d}\theta.
\end{align}
This integral is known in signal processing literature as the cross-correlation of $P_1$ and $P_2$. 
We have used $\theta_1$ as a basis for the differences, but we could have equivalently used $\theta_2$.
If the posterior distributions are normalized, the parameter difference distribution will also be normalized.
Even if the two distributions are not continuous at the prior boundary, the distribution of parameter differences is always continuous, although it might not be smooth.
In GW parameter estimation problems we typically have some periodic parameters. 
In these cases differences are defined up to a period but all above properties hold.

The prior distribution on parameter differences is obtained by calculating the cross-correlation of the two prior distributions.
If the prior distributions are the same then this distribution always peaks at zero parameter difference.

We can use the distribution of parameter differences to understand if parameter determinations from two datasets are in agreement or not.
Intuitively, if $P(\Delta \theta)$ had most of its support located at large values of $\Delta \theta$ then the two parameter sets are not compatible. 
In this case, we would call it a tension between the two events in parameter space.
In the lensing context this would mean that the two events are not lensed images of a single binary merger.

We can quantify the probability that there is a parameter shift by calculating the integral:
\begin{align} \label{Eq:ParamShiftProbability}
\Delta \equiv \int_{P(\Delta \theta)>P(0)} P(\Delta \theta) \text{d}\Delta \theta\,
\end{align}
that quantifies the posterior probability above the iso-density contour of no parameter shift, $\Delta \theta=0$~\cite{2012arXiv1210.1066T}.

While this allows us to calculate the statistical significance of a tension when distributions are non-Gaussian, we have the following conceptual problem. 
The result is not invariant under non-linear one-to-one changes of parameter bases/coordinates $\Delta\theta\rightarrow \Delta\tilde \theta$.
While the probability element $P(\Delta \theta) \text{d}\Delta \theta$ is invariant under such a change,  the probability density $P(\Delta \theta)$ is not and neither is the integration volume defined by its iso-contours:
\begin{align} \label{Eq:ParamShiftProb_transform}
& P(\Delta\theta) > P(\Delta\theta=0) \nonumber\\
& \hspace{0.1cm} \Rightarrow \hspace{0.1cm}
P(\Delta \tilde{\theta})\left| \det\frac{\text{d}\Delta \tilde{\theta}}{\text{d}\Delta \theta}\right|>P(\tilde{0})\left| \det \frac{\text{d}\Delta \tilde{\theta}}{\text{d}\Delta \theta}\right|_{0}
\end{align}
when the Jacobian determinant depends on parameters. 
This problem can be solved by considering a threshold that restores parameter invariance.   We choose the closely related likelihood threshold:
\begin{align} \label{Eq:ParamShiftProb_logLike_threshold}
\Delta _{\mathcal{L}} &\equiv \int_{\mathcal{L}(\Delta \theta)> \mathcal{L}(0)}P(\Delta \theta) \text{d}\Delta \theta\
\end{align}
which is fully parameter invariant.
In practice we calculate the likelihood threshold as $\log P(\Delta\theta) - \log \Pi(\Delta\theta) \geq \log P(0) - \log \Pi(0)$ since we have only access to posterior and prior flows trained on the outcomes of parameter estimation.

In cases where the two posterior distributions have the same prior $\Delta _{\mathcal{L}}$ will typically give higher values of shift probability with respect to $\Delta$.
This holds because the prior distribution on parameter differences, obtained calculating the difference distribution between the priors, would be peaked at zero, thus removing its contribution would lower threshold of zero shift and enhance the probability to exceed zero shift.
In the opposite case, when the prior distributions of the two events are different, $\Delta _{\mathcal{L}}$ will give lower values of the probability to exceed zero.

As outlined in~\cite{Raveri:2021wfz} the calculation of parameter shift probability proceeds as follows.
First, we calculate the distribution of parameter differences. 
This can be done with existing samples from the probability distributions of individual datasets and taking differences between corresponding samples. 
These differences constitute samples from the parameter difference distribution. 
Given $n$ samples from each posterior, we can generate up to $n^2$ samples for the difference distribution. 
However, generating and storing such a large number of samples is computationally expensive and typically unnecessary. 
Instead, we retain a smaller, representative subset of the $n^2$ samples. 
This simplification is valid because the parameter difference distribution, being a cross-correlation integral of individual distributions, is smoother and simpler than the original distributions. 
Consequently, retaining $O(n)$ samples is sufficient to accurately capture the structure of the distribution.
The same methodology applies to the prior distribution, using its samples.

The calculation of Eq.~\eqref{Eq:ParamShiftProb_logLike_threshold} is more involved.
A normalizing flow is trained on the samples of both the posterior and prior distributions, allowing us to compute efficiently probability and likelihood values.
The tension integral is then evaluated as a Monte Carlo integral. Specifically, for each sample of the difference distribution, the likelihood is evaluated using the trained normalizing flows for the posterior and prior. 
These likelihood values are compared to the likelihood value of zero shift $(\Delta\theta=0)$ and the fraction of samples above threshold gives the Monte Carlo estimate of Eq.~\eqref{Eq:ParamShiftProb_logLike_threshold}.

\subsection{Likelihood ratio test}\label{sec:likelihood.ratio}

In this section we discuss using the likelihood ratio as a measurement of agreement between GW events parameters.
When posteriors are Gaussian and parameters well measured this test is equivalent to parameter shifts.
In this context, we use a discrepancy between the two results as a warning sign of the presence of significant non-Gaussianities
and e.g.\ make lensing a bad fit to the data even in cases where a long but low probability density non-Gaussian tails in the parameter shift would otherwise favor the lensing hypothesis.

In our version of this test, unlike the usual maximum likelihood ratio version, we consider the ratio relative to the maximum posterior.
Our test statistic can then be written as:
\begin{align} \label{Eq:Like_ratio_test_def}
Q_\mathcal{L} &\equiv -2\left[\log{\mathcal{L}(\Delta\theta_{\rm f})}-\log{\mathcal{L}(\Delta\theta_{\rm MAP})}\right] 
\end{align}
where $\Delta\theta_{\rm f}$ is a fixed parameter difference and $\Delta\theta_{\rm MAP}$ is the parameter difference  at maximum posterior.
Note that in our application the fixed point is set to $\Delta\theta_{\rm f}=0$.
In this sense, our statistic quantifies how strongly the data favor parameter shifts. Specifically, a large value of the likelihood ratio indicates that the maximum posterior value \(\Delta\theta_{\rm MAP}\) is significantly more probable than \(\Delta\theta = 0\), or equivalently, that \(\Delta\theta = 0\) is strongly disfavored. Conversely, a likelihood ratio close to zero suggests that \(\Delta\theta_{\rm MAP}\) is not substantially preferred over \(\Delta\theta = 0\). Thus, this test effectively measures how unlikely the hypothesis \(\Delta\theta = 0\) is relative to the best-fit parameter difference according to the data.

Note the difference between this and usual likelihood ratio hypothesis testing. In this case we are considering the MAP and not the true maximum likelihood. 
The difference between these two points is that, if the prior is informative, the fit will not be free to completely optimize parameter degrees of freedom.
The advantage of our approach is that the maximum posterior can be easily calculated and retains a close connection to the result of parameter estimation. The maximum likelihood in contrast is not always well defined, as in the case of priors constrained by physical requirements that cannot be ignored, and might end up being in regions of parameter space that are not explored in parameter estimation, for example the value of $\chi_{\text{eff}}$ being between $-1$ and $1$.

In the context of GW parameter estimation this is even more relevant since the default prior is never flat - for example when considering parameters that describe angles.

This difference complicates significantly the calculation of the distribution of $Q_\mathcal{L}$ that we discuss in App.~\ref{App:likelihood.ratio.Proofs}. 
Assuming the Gaussian Linear Model~\cite{Raveri:2018wln} we can show that we can approximate the distribution as:
\begin{align} \label{Eq:LikelihoodRatioStatistic}
Q_\mathcal{L} \sim \chi^2(N_{\rm eff})
\end{align}
where $N_{\rm eff}$ is the number of effective parameters that a dataset is constraining over the prior, as defined in Eq.~\eqref{Eq:Neff}.
Here this quantifies the number of degrees of freedom of the test.
This result extends Wilks theorem~\cite{Wilks:1938dza} to the maximum posterior point rather than the maximum likelihood.

Note that, when we calculate the likelihood ratio we do not assume Gaussianity of the parameter posterior.
It is only when we try to associate a statistical significance to its value that we have to resort to assuming that its distribution derives from Gaussian posteriors.
As such, we treat this statistic as intermediate between Gaussian and fully non-Gaussian parameter shifts.

In practice we compute Eq.~\eqref{Eq:Like_ratio_test_def} from the normalizing flow model of each posterior that we use.
The evaluation speed and differentiability of the normalizing flow model allow us to use computationally intensive global optimization algorithms that require knowledge of the Jacobian, resulting in a stable identification of the maximum.
As discussed in~\cite{Raveri:2024dph} to avoid local maxima, we maximize the posterior by initializing the optimization from multiple points corresponding to the highest posterior values found in the parameter estimation chains.

\subsection{Reporting statistical significance} \label{sec:statistical.significance}

We always report probability results in units of effective standard deviations. 
Given an event with probability $p$, the number of effective standard deviations is defined as:
\begin{align}\label{Eq:EffectiveSigmas}
n_\sigma^{\rm eff}(p) \equiv \sqrt{2}\,{\rm Erf}^{-1}(p)\,,
\end{align}
where $n_\sigma^{\rm eff}$ represents the number of standard deviations that an event with the same probability would correspond to if drawn from a Gaussian distribution.  
In the following we always calculate $p$ as the probability of lensing consistency so that $n_\sigma$ is large when inconsistent.
Note that Eq.~\eqref{Eq:EffectiveSigmas} does not enforce Gaussianity of the distribution of $p$ but only matches it to its Gaussian corresponding value.
This scale is useful for examining the tails of the distribution of $p$, where highly statistically significant events reside, as it effectively acts as a logarithmic scale for probabilities.

\section{Gravitational wave catalogs}\label{sec:datasets}

In our work we consider different simulated and real data catalogs of GWs.
Our goal is to test our methods in controlled setups, as well as actual data sets.  
In all the cases, we perform the same parameter estimation pipeline that is applied to real data in LVK catalogs using \texttt{bilby}~\cite{Ashton:2018jfp}. 
We characterize the detectors by their power spectral density, which can be found in the public LIGO Document T2000012-v1~\cite{website:ligonoise}.

\subsection{Noise-varying events}\label{sec:data.noise}

In order to study the statistical properties of the parameter shift and likelihood ratio measures of lensing consistency, we begin with identical source injections that differ only in their noise realizations. 
We create two sets of 10 noise realizations for the same signal, named NV-1 and NV-2 for ``noise-varying". 
We study each set independently. 

The properties of the NV set are summarized in the first rows of Tab.~\ref{table:injections_parameters}.  
The two sets differ in their arrival time, which lags half a day between the two, shifting the detector positions relative to the source event and serves the purpose of testing changes in the detector basis.  
All the noise-varying simulations are analyzed with the current three detector network (HLV) at its design sensitivity. 
We choose their luminosity distance to be quite close, $\sim300$Mpc, so that these are very high network SNR. 
NV-1 has $\rho_\mathrm{ntw}\sim80$, while NV-2 has $\rho_\mathrm{ntw}\sim100$ due to the different detector orientations.

Since each event in a given set differs only by its noise realization, they all represent type-0 images of the same underlying event. Each of the 45 event pairs within that set is then expected to be consistent with lensing and analysed as such.

\begin{table*}
\centering
\begin{tabular}{lccccccccccccccccc}
\hline
\hline
Injection & $m_1\,[M_\odot]$ & $m_2\,[M_\odot]$ & $a_1$ & $a_2$ & $\phi_1$ & $\phi_2$ & $\phi_{12}$ & $\phi_{JL}$ & $\theta_\JN$ & ra & dec & $\psi$ & $\phi_\text{ref}$ & $d_L\,$[Mpc] & $t_\text{ref}\,$[sec] & Morse phase & Network\\
\hline
\hline
NV-1 & 30.4 & 27.7 & 0.2 & 0.6 & 1.5 & 2.3 & 3.0 & 0.93 & 2.41 & 3.24 & 1.37 & 1.1 & 2.43 & 311.7 & -0.02 & 0 & HLV\\
NV-2 & 30.4 & 27.7 & 0.2 & 0.6 & 1.5 & 2.3 & 3.0 & 0.93 & 2.41 & 3.24 & 1.37 & 1.1 & 2.43 & 311.7 & 43200 & 0 & HLV\\
GW01 & 35.6 & 30.6 & 0.2 & 0.1 & 0.6 & 0.3 & 1.2 & 0.5 & 0.8 & 1.0 & 0.52 & 0.7 & 2.0 & 1000.0 & 0.0 & 0 & HL\\
GW02 & 35.6 & 30.6 & 0.2 & 0.1 & 0.6 & 0.3 & 1.2 & 0.5 & 0.8 & 1.0 & 0.52 & 0.7 & 2.0 & 1900.0 & 0.0 & 0 & HL\\
GW03 & 35.6 & 30.6 & 0.2 & 0.1 & 0.6 & 0.3 & 1.2 & 0.5 & 0.8 & 1.0 & 0.52 & 0.7 & 2.0 & 1000.0 & 0.0 & 0 & HLV\\
GW04 & 35.6 & 30.6 & 0.2 & 0.1 & 0.6 & 0.3 & 1.2 & 0.5 & 0.8 & 1.0 & 0.52 & 0.7 & 2.0 & 1900.0 & 0.0 & 0 & HLV\\
GW05 & 35.6 & 30.6 & 0.2 & 0.1 & 0.6 & 0.3 & 1.2 & 0.5 & 0.8 & 1.0 & 0.52 & 0.7 & 2.0 & 1394.0 & 517988 & $\pi/2$ & HL\\
GW06 & 35.6 & 30.6 & 0.2 & 0.1 & 0.6 & 0.3 & 1.2 & 0.5 & 0.8 & 1.0 & 0.52 & 0.7 & 2.0 & 2650.0 & 517988 & $\pi/2$ & HL\\
GW07 & 35.6 & 30.6 & 0.2 & 0.1 & 0.6 & 0.3 & 1.2 & 0.5 & 0.8 & 1.0 & 0.52 & 0.7 & 2.0 & 1394.0 & 517988 & $\pi/2$ & HLV\\
GW08 & 35.6 & 30.6 & 0.2 & 0.1 & 0.6 & 0.3 & 1.2 & 0.5 & 0.8 & 1.0 & 0.52 & 0.7 & 2.0 & 2650.0 & 517988 & $\pi/2$ & HLV\\
GW09 & 35.6 & 30.6 & 0.2 & 0.1 & 0.6 & 0.3 & 1.2 & 0.5 & 0.8 & 0.3 & 0.18 & 0.5 & 3.77 & 1526.0 & 3451153 & 0 & HL \\
GW10 & 35.6 & 30.6 & 0.2 & 0.1 & 0.6 & 0.3 & 1.2 & 0.5 & 0.8 & 0.3 & 0.18 & 0.5 & 3.77 & 2900.0 & 3451153 & 0 & HL \\
GW11 & 35.6 & 30.6 & 0.2 & 0.1 & 0.6 & 0.3 & 1.2 & 0.5 & 0.8 & 0.3 & 0.18 & 0.5 & 3.77 & 1526.0 & 3451153 & 0 & HLV \\
GW12 & 35.6 & 30.6 & 0.2 & 0.1 & 0.6 & 0.3 & 1.2 & 0.5 & 0.8 & 0.3 & 0.18 & 0.5 & 3.77 & 2900.0 & 3451153 & 0 & HLV \\
GW13 & 15.4 & 12.6 & 0.2 & 0.1 & 0.6 & 0.3 & 1.2 & 0.5 & 0.8 & 1.0 & 0.52 & 0.7 & 2.0 & 433.16 & 79121.0 & 0 & HL\\
GW14 & 15.4 & 12.6 & 0.2 & 0.1 & 0.6 & 0.3 & 1.2 & 0.5 & 0.8 & 1.0 & 0.52 & 0.7 & 2.0 & 823.0 & 79121.0 & 0 & HL\\
GW15 & 15.4 & 12.6 & 0.2 & 0.1 & 0.6 & 0.3 & 1.2 & 0.5 & 0.8 & 1.0 & 0.52 & 0.7 & 2.0 & 433.16 & 79121.0 & 0 & HLV\\
GW16 & 15.4 & 12.6 & 0.2 & 0.1 & 0.6 & 0.3 & 1.2 & 0.5 & 0.8 & 1.0 & 0.52 & 0.7 & 2.0 & 823.0 & 79121.0 & 0 & HLV\\
GW17 & 15.4 & 12.6 & 0.2 & 0.1 & 0.6 & 0.3 & 1.2 & 0.5 & 0.8 & 1.0 & 0.52 & 0.7 & 2.0 & 603.82 & 428696 & $\pi/2$ & HL\\
GW18 & 15.4 & 12.6 & 0.2 & 0.1 & 0.6 & 0.3 & 1.2 & 0.5 & 0.8 & 1.0 & 0.52 & 0.7 & 2.0 & 1147.87 & 428696 & $\pi/2$ & HL \\
GW19 & 15.4 & 12.6 & 0.2 & 0.1 & 0.6 & 0.3 & 1.2 & 0.5 & 0.8 & 1.0 & 0.52 & 0.7 & 2.0 & 603.82 & 428696 & $\pi/2$ & HLV\\
GW20 & 15.4 & 12.6 & 0.2 & 0.1 & 0.6 & 0.3 & 1.2 & 0.5 & 0.8 & 1.0 & 0.52 & 0.7 & 2.0 & 1147.87 & 428696 & $\pi/2$ & HLV \\
GW21 & 15.4 & 12.6 & 0.2 & 0.1 & 0.6 & 0.3 & 1.2 & 0.5 & 0.8 & 1.41 & 0.54 & 0.5 & 3.77 & 661.0 & 4918029 & 0 & HL \\
GW22 & 15.4 & 12.6 & 0.2 & 0.1 & 0.6 & 0.3 & 1.2 & 0.5 & 0.8 & 1.41 & 0.54 & 0.5 & 3.77 & 1256.16 & 4918029 & 0 & HL \\
GW23 & 15.4 & 12.6 & 0.2 & 0.1 & 0.6 & 0.3 & 1.2 & 0.5 & 0.8 & 1.41 & 0.54 & 0.5 & 3.77 & 661.0 & 4918029 & 0 & HLV \\
GW24 & 15.4 & 12.6 & 0.2 & 0.1 & 0.6 & 0.3 & 1.2 & 0.5 & 0.8 & 1.41 & 0.54 & 0.5 & 3.77 & 1256.16 & 4918029 & 0 & HLV \\
\hline
\end{tabular}
\caption{\label{table:injections_parameters}
Summary of the physical parameters for the injections described in Sec.\ \ref{sec:datasets}: detector frame component masses ($m_{1,2}$), dimensionless spin magnitudes ($a_{1,2}$), the azimuthal angle of the spin vectors ($\phi_{1,2}$), the difference between the azimuthal angles of the individual spin vectors ($\phi_{12}$), the difference between total and orbital angular momentum azimuthal angles ($\phi_{JL}$), the angle between the total angular momentum and the line of sight ($\theta_\JN$), right ascension (ra), declination (dec), polarization angle ($\psi$), phase at reference frequency ($\phi_\text{ref}$), luminosity distance ($d_L$) and reference time ($t_\text{ref}$). The reference frequency is at $20$Hz. Injections with Morse phase equal to 0, $\pi/2$ and $\pi$ correspond to type I, II and III images respectively. 
}
\end{table*}

\begin{figure*}[!ht]
\centering
\includegraphics[width=\textwidth]{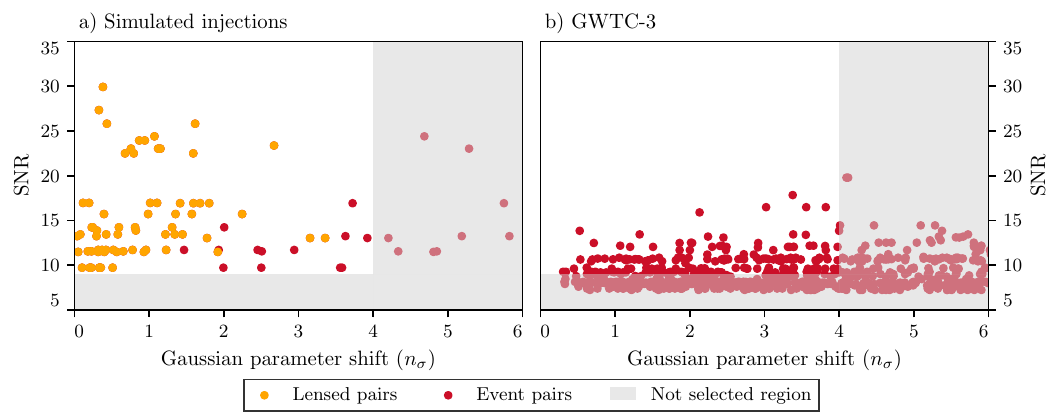}
\caption{\label{fig:sims_phazap_selected_snr}
Preselection of pairs with {\tt phazap}: SNR vs Gaussian parameter shift significance for pairs of the GW injection catalog, where the lensing status in known, in panel (a) and for pairs of the real events from GWTC-3, in panel (b) where all event pairs are shown.
Since injections are to designed to have network SNR in the range $12-30$, none of the simulated pairs are excluded by the SNR threshold. Selection is therefore driven solely by the significance of the Gaussian shift.  Notably, all lensed image pairs satisfy both thresholds and proceed to the full analysis. 
}
\end{figure*}

\subsection{GW injection catalog} \label{sec:data.simulations}

Following  Ref.~\cite{Ezquiaga:2023xfe}, we build a catalog of injections from triplets that contain a reference GW event, an image of this event (demagnified and phase shifted by $\pi/2$) and an unrelated event that is not a lensed image, but shares some parameters so as to provide interesting challenges for lensing identification.  
The lensed image of the reference event only changes its luminosity distance (due to the de-magnification), arrival time (due to the time delay), and Morse phase (due to the phase shift). 
We consider  four sets of these triplet events for a binary black hole system with chirp mass $\sim 25M_\odot$ (GW01 to GW12), and another four for a chirp mass $\sim 12M_\odot$ (GW13 to GW24). 
We keep the spins and inclination angle equal in all the cases. 
These four sets contain two luminosity distances so that there is a low-SNR configuration ($\rho_\text{ntw}\sim12-16$) and a high-SNR configuration ($\rho_\text{ntw}\sim22-30$) across the detector network.
We further take two detector configurations: a LIGO-only two detector network (HL) and a LIGO-Virgo three detector network (HLV). 
In total, we have 24 injections or 276 pairs.
We use the projected sensitivities of LIGO and Virgo for the fourth observing run, see Ref.~\cite{website:ligonoise} for details. 
The parameters of this set of simulations are summarized in Table \ref{table:injections_parameters}. 
It is to be noted that the parameters and configuration set ups of 15 of the simulated events (GW01-12, 16, 20, 24) coincide with those of Ref.~\cite{Ezquiaga:2023xfe}. 
However, the simulated events considered here correspond to a different noise realization.

\subsection{LIGO-Virgo-KAGRA GWTC-3}\label{sec:data.real}

We analyze the latest public LVK Gravitational Wave Transient Catalog, GWTC-3~\cite{LIGOScientific:2021djp} for binary black hole events without assuming any lower bound limit on the probability of having astrophysical origin ($p_\text{astro}>0$), and a network SNR $>9$, for a total of 86 events or 3655 pairs.

Real events differ from our simulations mostly by their lower SNR, which is due to the lower sensitivity of the detectors during the first three observing runs. 
Moreover, their parameters span a wider range of masses, spins, phases and sky localizations.

\subsection{Preselected pairs}\label{subsec:phazap_selection}

Within these event catalogs, we next select promising pairs for lensing candidates.
As the size of GW catalogs grows, it becomes increasingly computationally expensive to test all pairs for the compatibility of parameters.
For $n$ events, the  computational effort of a pairwise strong lensing search scales as $n^2$, which easily becomes prohibitive, even with highly optimized flow training and the statistical calculations we have discussed.

In addition, while in principle the non-Gaussian methods we have discussed could better quantify how excluded the lensing hypothesis is for the bulk of highly inconsistent pairs, in practice  Monte-Carlo integration and machine precision would limit its use those under $\sim 6\sigma$.
We therefore implement a fast preselection step that excludes pairs that are strongly inconsistent with the lensing hypothesis.

This fast selection step in our analysis relies on {\tt phazap} to pre-analyze pairs in each catalog~\cite{Ezquiaga:2023xfe}. 
This method estimates the lensing consistency between posteriors of pairs of events by computing mean parameter shifts in the detector basis defined in Sec.~\ref{sec:lensing} and interpreting them under a Gaussian approximation. 
We refer to this method as the ``Gaussian shift".

More specifically, if the posterior densities in Eq.~\eqref{Eq:param.diff.pdf} are Gaussian,
we can define the parameter shift distance as 
\begin{align}\label{eq:gaussian_distance}
    D_{12}\equiv D(\theta_1,\theta_2) = \sqrt{\Delta \theta^T (C_1+C_2)^{-1} \Delta \theta} \,,
\end{align}
where $C_1$ and $C_2$ are the covariances of each individual event posterior and the pair $\theta_1,\theta_2$ are the parameter means.
Covariances are computed wrapping periodic parameters so that the mean of the distribution is centered in the period.
Means are computed in the reference frame of the first event and in general $D_{12} \neq D_{21}$ because of reference frame choice.
As discussed in Sec.~\ref{sec:lensing}, the detector basis only optimally compresses information for an event in its own frame and comparing events where the detectors are oriented in different directions will always be less efficient.  
Among the two distances we select the one that excludes lensing to higher significance since a poor choice of reference frame can degrade the significance of rejection but does not in itself cause a false rejection of the lensing hypothesis.   
A true lensed pair will exhibit consistent parameters in both orderings.
Strong lensing introduces a constant phase shift to the detector phases, so we calculate the distance for all possible phase shifts.  In this case we take the minimum distance since any of these possibilities is consistent with lensing.

Since this method assumes $\Delta \theta$ to be Gaussian-distributed, we can compute statistical significance assuming its distribution to be:
\begin{align} \label{Eq:gaussian_distance_statistic}
 D_{12}^2 \sim \chi^2(\Ngs) \,,
\end{align}
where $\Ngs$ is the number of effective constrained parameters under the Gaussian shift (``gs") approximation.
Specifically,
in accordance with Ref.~\cite{Ezquiaga:2023xfe}, we account for the trials over the Morse phase by taking the prior for phase parameters to be $\mathcal{C}_\Pi = \delta_{d_1,d_2}(\pi/2)^2/12 $ for $d_1,d_2 \in\mathrm{H,L,V}$.  
For counting these three parameters, we then take Eq.~\eqref{Eq:Neff} to compute $N_{\phi_{\rm d}}$
and add them to the separate computation of Eq.~\eqref{Eq:Neff} for $N_{\rm other}$ for the three other detector parameters, $\tau_{\mathrm{HL}},\tau_{\mathrm{LV}},\Delta \phi_f$, i.e.\ $\Ngs=N_{\phi_{\rm d}}+N_{\rm other}$ neglecting the covariance between the two sets.   
This also differs from the {\tt phazap} parameter counting in~\cite{Ezquiaga:2023xfe} where to be conservative, their effective number of constrained parameters was defined as $N_{\rm eff} = N_{\phi_{\rm d}}+3$.
Consequently, $n_\sigma$ for the Gaussian shift in the {\tt phazap} selection here is larger than it would be if converted from the $p$-values given by the code {\tt phazap} itself. 
Note that both the Gaussian shift and {\tt phazap}  parameter counting differs from $N_{\rm eff}$ defined by  Eq.~\eqref{Eq:Neff} with the  covariance matrix and priors of the original parameter estimation. 
In particular, the $p$-value derived from $\Ngs$ accounts for the reduced shifts due to the trials over the Morse phase whereas neither the likelihood ratio nor the  non-Gaussian parameter shifts as reported do.

We conservatively select pairs with Gaussian shift disagreements of $n_\sigma\le 4$ and proceed with the analysis only for these pairs. Posterior distributions with disagreements exceeding this are deemed inconsistent with the lensing hypothesis and are therefore excluded. This threshold is set high enough to ensure that typical non-Gaussianities have should not lead to false negatives. 
We refer to this selection process as the \texttt{phazap} selection.

All of the NV pairs satisfy the \texttt{phazap} selection, while some pairs are not selected for the other two catalogs as shown in Fig.~\ref{fig:sims_phazap_selected_snr}(a) for the GW injection catalog and Fig.~\ref{fig:sims_phazap_selected_snr}(b) for GWTC-3.
For the former, all pairs satisfy the SNR cut, and the selection is driven solely by the \texttt{phazap} significance of the parameter difference. No lensed pair is excluded by the significance threshold of 4 sigma.
In total the full analysis is performed on 81 pairs out of the 276 in the catalog.

For the GWTC-3 pairs, as we can see in Fig.~\ref{fig:sims_phazap_selected_snr}(b), the SNR cut is more stringent but the significance cut still allows for a large number of pairs to be selected. The full analysis is then only performed on 185 pairs out of the 3655 in the catalog.  This substantially reduces the computational costs of the GWTC-3 analysis.

\begin{figure*}[!ht]
\centering
\includegraphics[width=\textwidth]{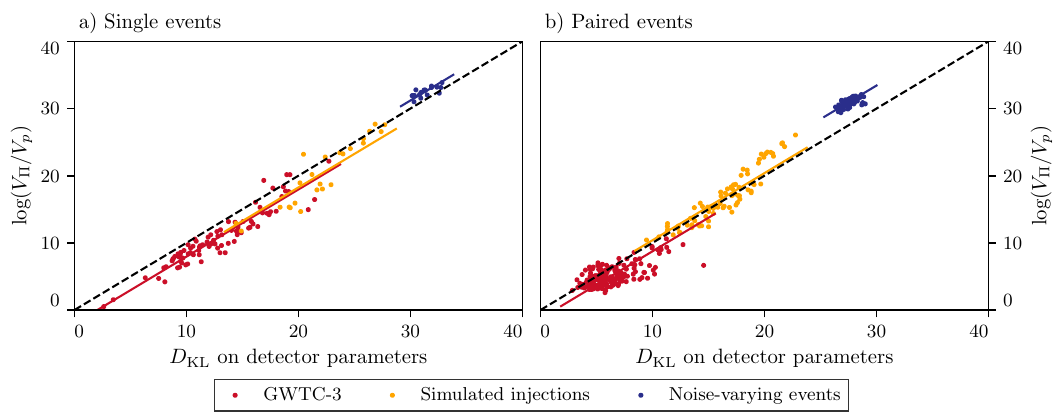}
\caption{\label{fig:info.volume.single.detector}
Relation between KL divergence and posterior/prior volume ratio computed in the detector parameter basis for individual events, in panel (a), and for events pairs, in panel (b). 
Each point corresponds to a single/pair event from different catalogs, as shown in the legend. 
The dashed line corresponds to linear trend. Monte-Carlo error estimates are smaller than the markers. 
The linear trend is in good agreement with Eq.~\eqref{Eq:avg.KL.volume}.
}
\end{figure*}

\section{Choosing an Informative Basis}\label{sec:results.info}

In this section we discuss how to choose an informative parameter basis for lensing searches based on the information content that it captures.
In Sec.~\ref{sec:infoandvolume} we validate the relationship between information and parameter volume for single events and events pairs.
In Sec.~\ref{sec:information.parameter.bases} we discuss the specific parameter bases that are most informative for lensing searches.

\subsection{Information and volume}\label{sec:infoandvolume}

\begin{figure*}[!ht]
\centering
\includegraphics[width=\textwidth]{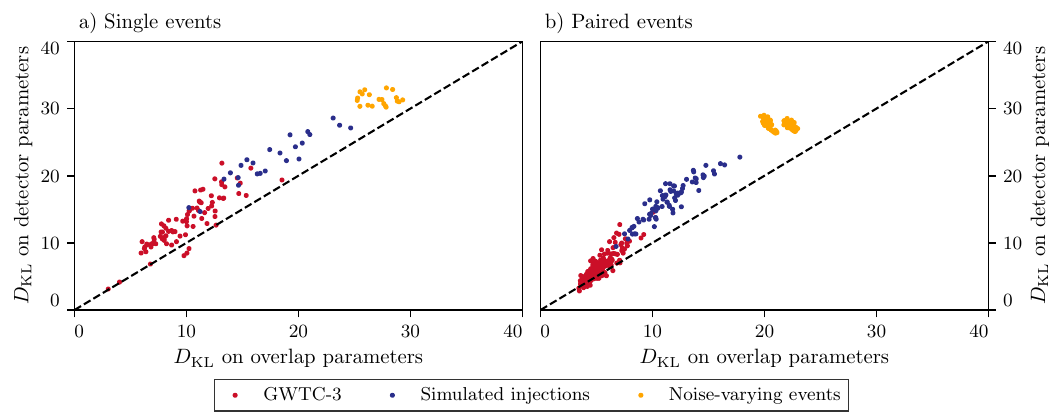}
\caption{\label{fig:2_KL_detector_vs_overlap_individual_events}
Comparison between KL divergences computed on the 9 overlap parameters and the 6 detector basis parameters. 
Panel (a) shows the KL divergence for individual events, while panel (b) shows the KL divergence for pairs of events.
Each point corresponds to a single event from different catalogs, as shown in the legend. 
The dashed line highlights equal information.
Monte-Carlo error estimates are smaller than the markers.  Despite compression into fewer parameters, the detector basis typically contains more information than the overlap basis.
}
\end{figure*}

We extract here the relationship between the KL divergence and posterior volume for gravitational wave events that we discussed in Sec.~\ref{sec:information.content}.
Fig.~\ref{fig:info.volume.single.detector}(a) compares the two  for detector basis parameters across individual events from the three catalogs analyzed. The figure includes all events from the catalogs. For this parameter basis, which includes three periodic parameters, we compute the covariance by centering the mean of the distribution.

The overall trend is well-aligned with expectations, showing a linear relationship with minimal deviations. 
Note that the linear relationship is expected in an ensemble-averaged sense as is best illustrated by the NV catalog. Noise realizations of the same event introduce scatter and the relationship is strictly valid only for perfectly Gaussian distributions in both priors and posteriors. 
While the prior distributions are not Gaussian, they are sufficiently similar across events in each catalog to contribute minimal variations, primarily manifesting as a uniform offset between catalogs in the plots. 
The posterior distributions, in contrast, is better approximated by a Gaussian as the SNR increases, especially between the real and simulated catalogs.

We fit a line with a fixed slope (equal to one) to the results for each catalog, observing that each population  aligns pretty closely with the fit, although showing slight variations in offset.
These small variations are expected since as the SNR changes, the relative amount of information coming from different parameters also changes, each with different non-Gaussianity in their priors and posteriors.  This is especially true of the three phase parameters which contribute more information at high SNR.

At very low SNR both $D_\KL$ and $\log (V_\Pi/V_p) \rightarrow 0$ but there are cases where the non-Gaussian information of the posterior that is  counted in the former but not the latter is important, whereas the converse is not true: if $D_\KL \rightarrow 0$ the whole posterior distribution approaches the prior and $V_\Pi/V_p \rightarrow 1$ by definition.

We then compute the KL divergence for the distribution of parameter differences in the detector basis, restricting the analysis to  the preselected  pairs in Sec.~\ref{subsec:phazap_selection}. 
Fig.~\ref{fig:info.volume.single.detector}(b) focuses on the detector basis parameters, again accounting for three periodic parameters by wrapping them to center their mean. 

As in the previous case this analysis shows a robust linear relationship.
Notably, when fitting each catalog with a linear relation, the offset trend seems more pronounced than in the analysis of single events. 
The derived parameter posteriors typically become more non-Gaussian when casting one event in the detector basis of the other.

Note that comparing the scale of the Fig.~\ref{fig:info.volume.single.detector}(a) and Fig.~\ref{fig:info.volume.single.detector}(b) we can appreciate the effect of marginalizing over parameter values when considering differences.
The KL divergence factorizes for parameters that are independent for both prior and posterior, as it happens in the joint parameter distribution, in Eq.~\eqref{Eq:Def_KL_div}. That distribution then has, for similar events, twice the information of the single event. When considering parameter differences we then marginalize over a copy of the parameters, lowering the overall information content to a level that is comparable with the one of the single event. 

\subsection{Information and parameter bases}\label{sec:information.parameter.bases}

Thanks to the data processing inequality of Eq.~\eqref{Eq:DPI}, we can quantify the efficiency of different parameter bases at capturing information on GW events.
We can think of the standard parameter estimation as a non-linear data compression: for GW events that match the templates of the waveform approximant  this compresses the waveform data to 15D, as we discussed in Sec.~\ref{sec:lensing}. For lensing searches, we further compress this parameter space, considering the overlap basis (9D) or the detector  basis (6D).
All these compressions will be to some extent lossy. The comparison of the KL divergences for each of them allows us to test which parameter basis retains more information.
We compute KL divergence from prior and posterior distributions of individual events as in Eq.~(\ref{Eq:MC_KL_div}).

In Fig.~\ref{fig:2_KL_detector_vs_overlap_individual_events}(a) we compare the information content, for each event, computed in the overlap and detector parameter space.
We can see that the KL divergence on the detector basis is higher than in the overlap basis for almost all events.
This shows that, despite having fewer parameters, the detector basis is more efficient at compressing the event's information.

We then investigate whether such result holds when considering parameter differences, which are most relevant to our lensing problem.
In Fig.~\ref{fig:2_KL_detector_vs_overlap_individual_events}(b) we compare the KL divergence computed from prior and posterior of the difference distributions of each pairs of events in the overlap basis against the KL divergence in the detector basis.
Also in this case we can see that detector basis retains more information with most points above the diagonal.

\begin{figure*}[!ht]
\centering
\includegraphics[width=\textwidth]{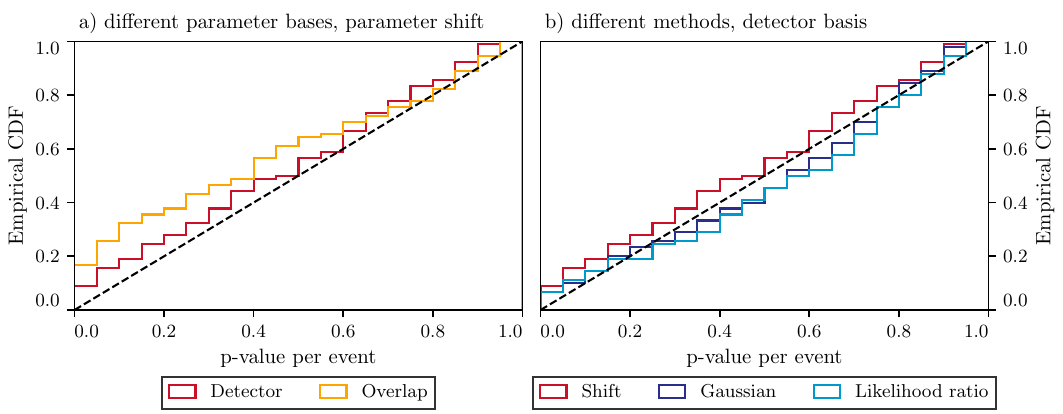}
\caption{ \label{fig:noisy_sims_bases_pop_cdf}
Empirical cumulative distribution functions (CDFs) of computed $p$-values for the parameter shift estimator applied to the NV catalog, using the two different parameter bases: Detector parameters and Overlap parameters. The black dashed line represents the ideal cumulative distribution function, CDF$[p]=p$, which corresponds to the expected distribution of $p$-values under the null hypothesis-that is, assuming that parameter shifts arise solely due to random noise fluctuations and not from any physical effect such as lensing. In such a case, the probability of observing a $p$-value less than any given threshold $p$ should be exactly $p$, resulting in a uniform distribution. The closer a curve lies to the diagonal, the better the agreement with this ideal behavior, revealing that the parameter shift in the detector basis performs the best.
}
\end{figure*}
%

\section{Identifying Lensing Candidates Pairs}\label{sec:results.identification}

Having established the detector basis as optimally informative, we proceed to analyze the different GW catalogs with our normalizing flow method mainly in this basis.
In Sec.~\ref{sec:results.noisy_sims} we apply the different methods to the NV catalog, which is composed of events that are consistent with lensing and differ only by noise realizations.
In Sec.~\ref{sec:results.sims} we show the results for the GW injection catalog, which consists of simulated lensed and non-lensed events with known parameters.
In Sec.~\ref{sec:results.sims.lensing_plane}, we use these studies to define an identification strategy involving parameter shifts and KL information.

\subsection{Noise-varying catalog} \label{sec:results.noisy_sims}
In this section we investigate how different techniques and different parameter bases impact the calculation of statistical significance of a parameter shift.
To do so we employ the two  ``NV'' catalog of events.  
In this case, all pairs that we can form out of the catalog are effectively type-0 images that are only distinguished by a given realization of the noise. The detector positions are the same between the events in each pair and this reduces the inefficiency of projecting one event onto the detector basis of the other as well as the non-Gaussianity that projection entails.

The difference in noise realizations will propagate to parameter estimation shifting each of the posterior distributions. 
This shift is expected and unavoidable. Therefore even though each pair in the NV catalogs is the same event, and thus also consistent with lensing, each method will report a significance or $p$-value for rejecting the lensing hypothesis that is not simply zero.
In fact we can use the NV catalogs to assess whether the distribution of reported $p$-values reflect the ensemble of noise realizations.  
For example, for the parameter shift estimator we can ask: do 50\% of pairs have $p > 0.5$? 
If we perform this test across all $p$-levels, we are effectively conducting a Kolmogorov-Smirnov (KS) test to check whether 
the empirical cumulative distribution function (CDF) of the $p$-values is consistent with the ideal CDF, i.e. CDF$[p]=p$.

In Fig.~\ref{fig:noisy_sims_bases_pop_cdf} (a) we show the empirical CDF of the $p$-values computed for the NV catalog using the two different parameter bases: detector and overlap.
As in the KS test, the closer a line is to the diagonal, the better the agreement between the empirical and expected distributions.

As we can see the parameter shift computed over detector parameters are fairly close to the diagonal.
The KS test performed comparing the empirical distribution to the ideal distribution gives a significance of 73\%, which indicates that the distribution of $p$-values is consistent with the expected uniform distribution.
On the other hand we can see that the distribution of the same quantity, when computed on overlap parameter space, gives a line that is further away from the diagonal, resulting in a KS test significance of 0.7\%, significantly lower than the previous one.
As we can see the overlap line grows faster at small p-values indicating a pile-up of pairs with small $p$-values, which is not expected from the noise realizations.

In Fig.~\ref{fig:noisy_sims_bases_pop_cdf} (b) we compare the performances of the different methods we have discussed in Sec.~\ref{sec:lensing} to compute the significance of a parameter shift in the detector basis.
The NV simulations are high enough SNR that the posterior distributions are close to Gaussian.
We then expect that the three methods would give similar results as they are equivalent in the limit of Gaussian posteriors.
As in the previous figure parameter shift is very close to the diagonal, with a KS test significance of 73\%.
Gaussian distance is also very close to the diagonal, with a KS test significance of 19\%.
The likelihood ratio test statistic is the furthest from the diagonal, with a KS test significance of 5\%.
Note that the KS test significance is very sensitive to small deviations from the ideal distribution, especially when the number of pairs is large, and it is customary to consider a conservative threshold, usually below 1\%, to flag potential issues.

These results overall validate the use of the three estimators we have discussed in Sec.~\ref{sec:lensing} to compute the significance of a parameter shift.
It also highlights the fact that this KS test is very sensitive to small deviations from the theoretical distribution as the number of pairs increases. In fact the lines for Gaussian distance and likelihood ratio are very close to each other.

\begin{figure*}[!ht]
\centering
\includegraphics[width=\textwidth]{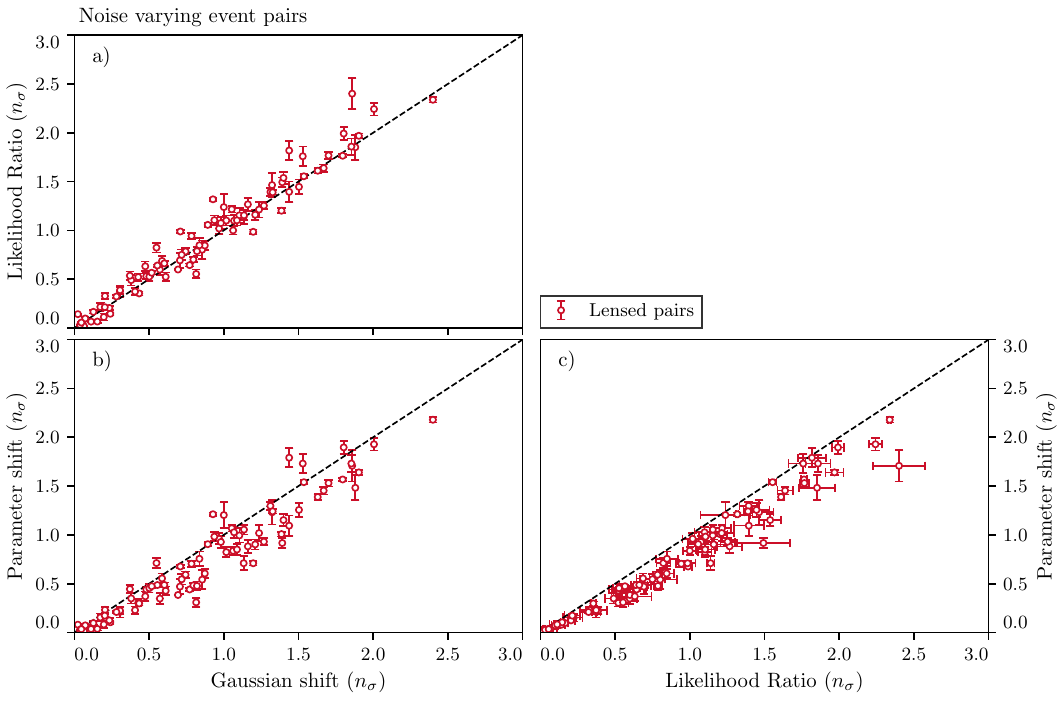}
\caption{\label{fig:NV_metrics}
Comparison between lensing metrics for noise simulations: Gaussian and non-Gaussian parameter shifts significance, likelihood ratio significance. Each point represents a different pair of simulated noise realizations. Error bars correspond to flow model variance and whereas for the Gaussian shift no error estimates are available. Note that for this catalog all pairs are type-0 lensed images.
}
\end{figure*}

This is further confirmed in Fig.~\ref{fig:NV_metrics}, where we compare the significance of the parameter shift and the likelihood ratio estimators in the detector basis.
As we can see in all panels, all methods give similar results, modulo a small offset towards higher significance for the likelihood ratio test statistic and the Gaussian shift.

When posteriors are Gaussian and parameters well measured this is the expected trend as the significance estimated with all three methods has to be the same.
Fig.~\ref{fig:NV_metrics} verifies this trend with the NV pairs, which are high-SNR and exhibit nearly Gaussian posterior distributions.

A discrepancy between the significance between the tests functions as a warning sign of the presence of significant non-Gaussianities, as we will discuss in more details for the analysis of both simulated and real pairs of events in the following sections.

\subsection{GW injection catalog} \label{sec:results.sims}

In this section we analyze the performances of different methods on simulated lensed and not-lensed injections.
In Sec.~\ref{sec:results.sims.differentparameters} we compare the significance of the parameter shift and likelihood ratio test statistics in the detector basis and overlap parameter bases.
In Sec.~\ref{sec:results.different_estimators} we comment on the significance of a lensing rejection using different methods, in the detector basis.

\subsubsection{Lensing significance in different parameter bases}\label{sec:results.sims.differentparameters}

\begin{figure*}[!ht]
\centering
\includegraphics[width=\textwidth]{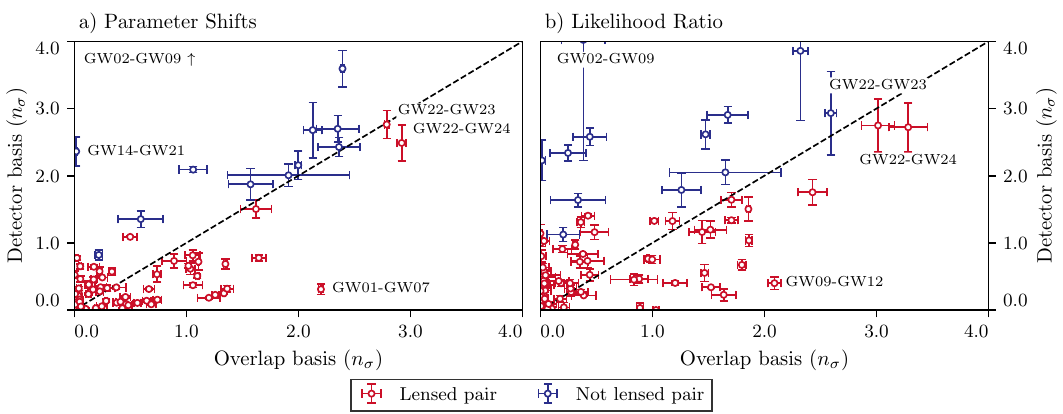}
\caption{\label{fig:4_param_shift_overlap_vs_detector_sims_like}
Comparison plot of non-Gaussian parameter shift, panel (a), and likelihood ratio significance, panel (b), computed in the overlap and detector parameter bases.
Each point represents a different pair of simulated events.
Error bars show flow variance.  Labeled pairs represent special cases discussed in the text and $\uparrow$ for {\bf GW02-GW09} indicates the pair is offscale.
}
\end{figure*}

We begin by comparing the significance of the parameter shift and likelihood ratio test statistics in the detector basis and overlap parameter bases, as shown in Fig.~\ref{fig:4_param_shift_overlap_vs_detector_sims_like}.

\begin{figure*}[!ht]
    \centering
    \begin{minipage}[t]{0.48\textwidth}
        \centering
        \includegraphics[width=\linewidth]{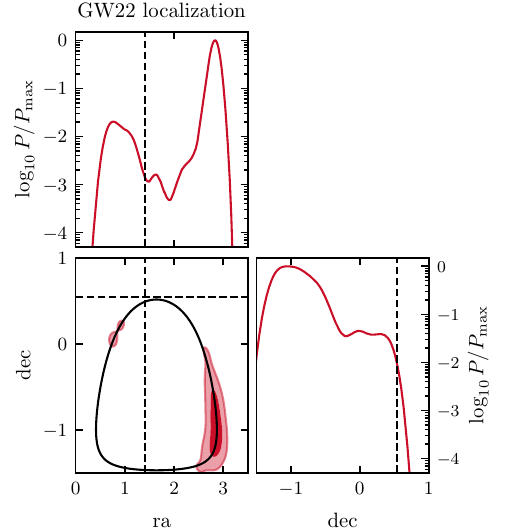}
        \caption{\label{fig:GW22_noise_realization_wrong_loc}
        Localization posterior distribution for \textbf{GW22} simulated event, involved in the false rejection of the lensing hypothesis with \textbf{GW23, GW24}. Dashed markers correspond to the true injected values for sky position parameters $\ra$, $\dec$, which is outside the 95\% credible regions of the posterior distribution.
        The continuous black line shows the best fit $\tau_{\rm HL}$ timing localization ring.
        }
    \end{minipage}
    \hfill
    \begin{minipage}[t]{0.48\textwidth}
        \centering
        \includegraphics[width=\linewidth]{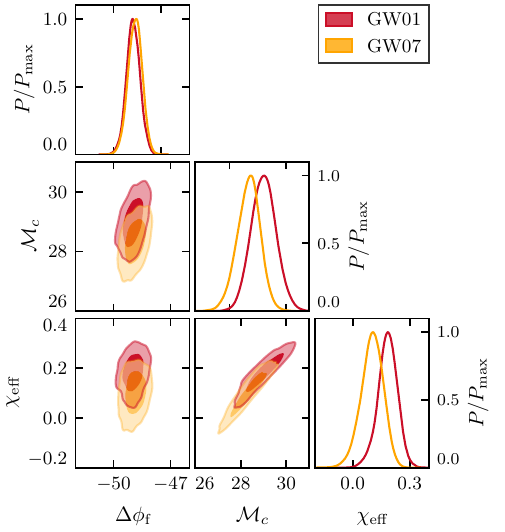}
        \caption{\label{fig:GW01_GW07_chirp_chieff_corr}
        Posterior distributions for the two injections \textbf{GW01} and \textbf{GW07}, a lensed pair. 
        Overlap parameter shifts disfavor lensing whereas detector basis shifts correctly report consistency due to the different representations of the mass parameters $\mathcal{M}_c$ vs $\Delta\phi_f$. The chirp mass is degenerate with the spin parameter $\chi_{\rm eff}$  and the discrepancy is along the weakly constrained direction whereas $\Delta\phi_f$ isolates the more robust, best constrained direction.
        }
    \end{minipage}
\end{figure*}

Comparing the significance of detector basis and overlap basis lensing results, we find that in both bases the majority of lensed pairs show strong consistency.
As we can also see, in the detector basis, for both methods, the significance of rejection of not lensed pairs is generally higher than in the overlap basis.
This suggests that the detector basis more effectively captures the significance of lensing consistency and is consistent with the fact that the detector basis is more efficient at capturing information, as we have shown in Sec.~\ref{sec:infoandvolume}.

In both plots we can notice two false negatives, i.e.\ two pairs that are actually lensed images but they exhibit strong inconsistency for both statistics and in both bases, with overlap being slightly more erroneous: \textbf{GW22-GW23} and \textbf{GW22-GW24}. 
We traced the difference to a chance noise fluctuation in the shared single event \textbf{GW22} which shifts the localization parameters away from agreement, as shown in Fig.~\ref{fig:GW22_noise_realization_wrong_loc}.
This example also highlights the fact that although events are independent, chance occurrences in a catalog of pairs can be more common since different pairs can share the same anomalous event.

\begin{figure}[!h]
\centering
\includegraphics[width=\columnwidth]{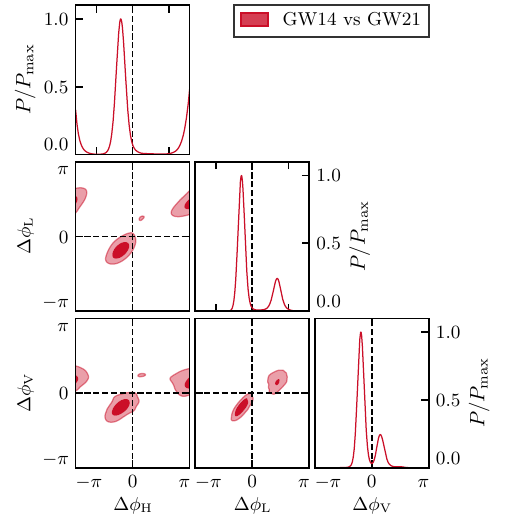}
\caption{\label{fig:GW14_vs_GW21_detector_phases_diff}
Parameter difference posterior distributions for injections \textbf{GW14} and \textbf{GW21}, a not lensed pair. Most of the ability to reject the lensing hypothesis comes from differences in the  $\phi_\dH, \phi_\dL,\phi_\dV$ phases which are inaccessible in the overlap basis.
}
\end{figure}

It is also instructive to examine the extreme cases where various lensing metrics disagree in Fig.\ \ref{fig:4_param_shift_overlap_vs_detector_sims_like}. These cases are:
    
\begin{itemize}
    \item \textbf{GW01-GW07}: This system consists of a lensed pair of the Type I and II images, where the former has been detected by two detectors (HL) with high SNR ($\rho_{\rm{ntw}} = 30$), while the latter has been observed by three detectors (HLV) with high SNR ($\rho_{\rm{ntw}} = 24$) (Table \ref{table:injections_parameters}). This pair shows agreement in  \( \Delta \phi_f \) but not in the chirp mass, leading to overlap basis to incorrectly disfavor the lensing hypothesis at $\sim 2\sigma$ in both parameter shift and likelihood ratio significance. 
    This discrepancy in masses seen in the overlap basis is clearly visible in the $\mathcal{M}_c-\chi_{\text{eff}}$ subplot in Fig.~\ref{fig:GW01_GW07_chirp_chieff_corr}. There is a noise fluctuation in the weakly constrained parameter combination but the best constrained direction is in agreement. 
    $\Delta \phi_f$ compresses this information into a single parameter which is not sensitive to the noise fluctuation. 
    \item \textbf{GW09-GW12}: This corresponds to a pair of Type 0 images from the same lensed source. The first one is seen by two detectors (HL) with low SNR ($\rho_{\rm{ntw}} \sim 17$), while the second event is observed by three detectors (HLV) and with high SNR ($\rho_{\rm{ntw}} \sim 23$) (Table \ref{table:injections_parameters}). This consists in a lensed pair, with exactly the same injected true parameters except for luminosity distance, to simulate different detector SNR configurations.
    It seems to exhibit the same features as \textbf{GW01-GW07}, showing agreement in  \( \Delta \phi_f \) but not in the chirp mass, leading to overlap basis to incorrectly reject the lensing hypothesis. This is again relatable to the mass-spin  degeneracy that we showed for the previous pair in Fig.~\ref{fig:GW01_GW07_chirp_chieff_corr}. The difference between the significance of parameter shift and likelihood ratio for this pair is due to strong non-Gaussianities in the two events posteriors, that were not present in \textbf{GW01-GW07}. In particular the $\Delta \theta_\JN$ parameter difference distribution is multimodal.
    \item \textbf{GW14-GW21}: This corresponds to a pair of not lensed events, both detected by LIGO only (HL), but the first is a `low-SNR' detection ($\rho_{\rm{ntw}} = 14$), while the second  is higher ($\rho_{\rm{ntw}} = 24$). The detector basis parameter shift disfavors lensing because of the  HL detector phases, as we can see in Fig.\ \ref{fig:GW14_vs_GW21_detector_phases_diff}.  Meanwhile, the overlap parameters are consistent with lensing because the overlap basis does not retain phase information.  
    Additionally, both injections correspond to two-detector events, leading to sky localization forming ring-like structures that intersect, despite the fact that these have entirely different sky locations. Rejection of the lensing hypothesis then rests mainly on the detector phases.
    \item \textbf{GW02-GW09}: This system consists of a pair of not lensed events, both have been detected by two detectors (HL) but the former has low SNR ($\rho_{\rm{ntw}} = 17$), while the latter has high SNR ($\rho_{\rm{ntw}} = 23$) (Table \ref{table:injections_parameters}). The significance of the rejection is higher in the detector basis because the overlap basis does not include phase parameters, so it is rejecting the lensing consistency just based on source localization parameters.
    This is simply another example of same issue seen in \textbf{GW14-GW21} but with different relative significances for the phases and localization. In Fig.~\ref{fig:4_param_shift_overlap_vs_detector_sims_like} this pair only explicitly appears in Panel (b), for the likelihood ratio, since it is rejected by $>4\sigma$ by parameter shifts in the detector basis.
\end{itemize}

Thus, even the outliers in these population plots reinforce the conclusion that detector parameters provide a more reliable means of identifying strongly lensed pairs while effectively rejecting non-lensed pairs.

\subsubsection{Lensing significance with different estimators}\label{sec:results.different_estimators}
\begin{figure*}[!ht]
\centering
\includegraphics[width=\textwidth]{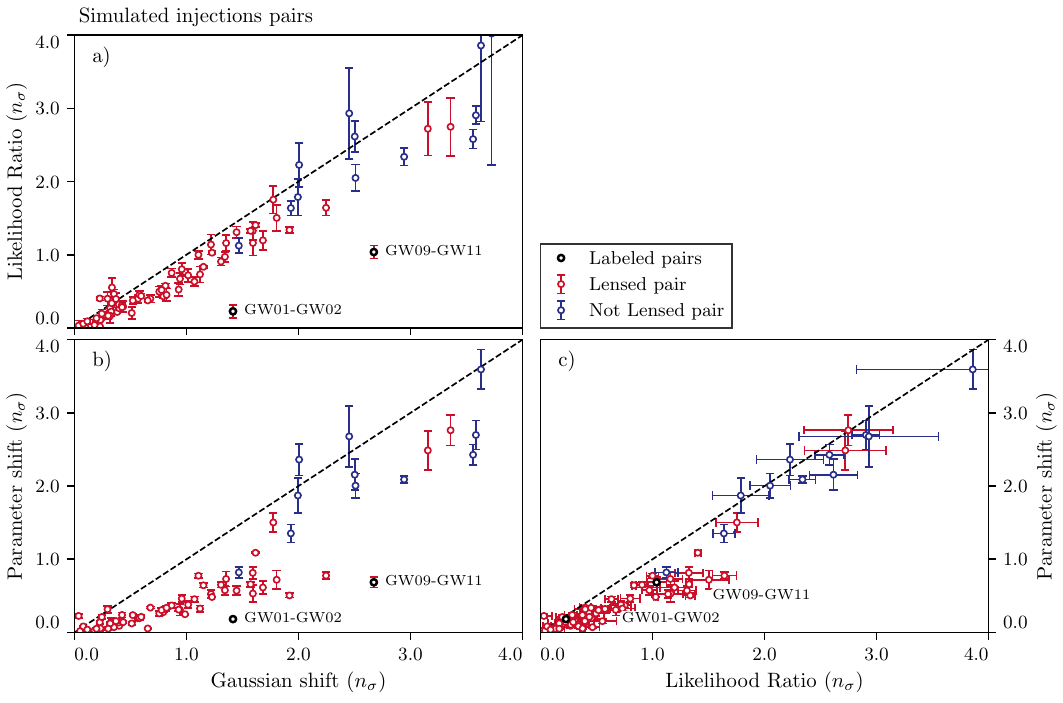}
\caption{\label{fig:tensiometer_like_ratio_smooth_vs_param_shift}
Comparison between lensing metrics for simulated injections: Gaussian and non-Gaussian parameter shifts significance, Likelihood ratio significance. 
Each point represents a different pair of simulated events. 
Error bars correspond to flow model variance whereas for the Gaussian shift no error estimates are available. 
Different colors correspond to either lensed and not lensed pairs, as shown in the legend.
Details of the labeled pairs are given in the main text.
}
\end{figure*}

In this section, we compare different estimators within the same parameter basis. 
In particular, based on the conclusions drawn in the previous section, we focus on results obtained in the detector basis only, and we show these results in Fig.~\ref{fig:tensiometer_like_ratio_smooth_vs_param_shift}.

For typical pairs  $n_\sigma$ is highly correlated  between the estimators but with a notable trend in the specific relationship. From  lowest to highest $n_\sigma$ we have: parameter shifts, likelihood ratio, and Gaussian shift.

This trend is expected since the non-Gaussian parameter shift allows to properly account for a slower decay of the tails of the difference distribution, which is not captured by the Gaussian shift or the likelihood ratio test statistic.  
Of course in the low $n_\sigma$ regime all three agree that the pairs are consistent with lensing, and disagree only on the probability for lensing that they assign. 

Furthermore, note that in this consistency regime, all these metrics report $p$-values for lensed pairs that are different from the expected uniform distribution, that we have shown in Sec.~\ref{sec:results.noisy_sims}. For example for parameter shifts, noticable more than 68\% of lensed pairs have $n_\sigma<1$.
This is expected since the catalog is lower SNR than the NV catalog and the prior becomes more informative.
Crucially the prior has no noise fluctuations and, in the limit of completely uninformative data, all three estimators should give the same $p$-value for all noise realizations. 

Therefore, as we shall discuss below, for pairs that are consistent with lensing, a more useful criteria for ranking is the KL divergence since parameter consistency with a high level of parameter information supplied by the data is a good indicator of lensing.

In the intermediate regime, from Fig.~\ref{fig:tensiometer_like_ratio_smooth_vs_param_shift} we can see that the parameter shift better separates lensed pairs from unlensed pairs with respect to the other two metrics which makes it useful as the primary metric.

In the high $n_\sigma$ regime where lensing is correctly rejected for not-lensed pairs, all three estimators  perform similarly for most pairs.   
Notably, the pairs lie closer to the diagonal of equal significance between the metrics.

\begin{figure}[!ht]
\centering
\includegraphics[width=\columnwidth]{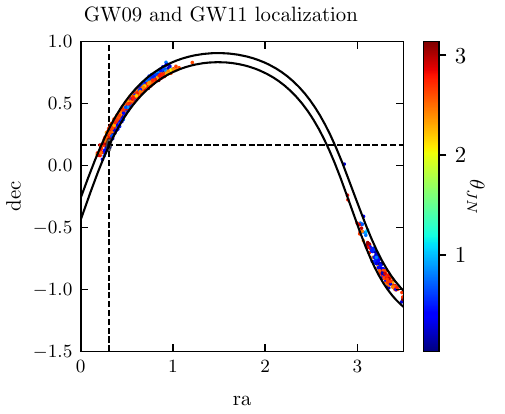}
\caption{\label{GW09_localization_inclination}
Localization (ra, dec) posterior distributions for injection \textbf{GW09} colored by inclination posterior values which shows crossed rings. \textbf{GW11}, a lensed copy, is so well localized that we use the vertical and horizontal dashed lines to highlight its true position which is in agreement with only one of the \textbf{GW09} arcs, leading to problems with the Gaussian shift statistic. The black continuous lines represent the 2.5$\sigma$ region of the time delay constraint on $\tau_{\rm HL}$.
}
\end{figure}

Aside from these trends that the majority of pairs follow, there are outliers in the differences between the various metrics in 
Fig.\ref{fig:tensiometer_like_ratio_smooth_vs_param_shift} which indicate specific problems in the techniques.
These cases are:

\begin{itemize}
    \item \textbf{GW09-GW11} This system consists of a lensed pair of Type 0 images, where the lensing hypothesis is disfavored by the Gaussian shift ($\sim 2.7\sigma$) but allowed by both parameter shift and likelihood ratio. 
    The pair also have the same `high' SNR configuration ($\rho_{\rm{ntw}} \sim 23$), but GW09 has been detected by two detectors (HL) while GW11 has been observed by three detectors (HLV) (Table~\ref{table:injections_parameters}).
    As we can see from Fig.~\ref{GW09_localization_inclination}, the localization of GW09 splits into two intersecting arcs, representing two modes of the inclination $\theta_\JN$, only one of which intersects the true location, but both within the 2.5$\sigma$ region of the timing constraint on $\tau_{\rm HL}$ which limits the extent of the arcs.  
    The Gaussianization of this multimodal localization exacerbates the significance of the fluctuation in $\tau_{\rm HL}$, whereas parameter shifts and likelihood ratios correctly account for it.
    \item \textbf{GW01-GW02} This system consists of a lensed pair of Type 0 images, where the lensing hypothesis is somewhat disfavored by the Gaussian shift ($\sim 1.4\sigma$) but allowed by both parameter shift  and likelihood ratio.
    Both events have been detected only by LIGO (HL), but while GW01 is in the `high' SNR configuration ($\rho_{\rm{ntw}} \sim 30$), GW02 is in the `low' SNR configuration ($\rho_{\rm{ntw}} \sim 17$) (Table \ref{table:injections_parameters}). We find that the reason for the discrepancy is the same as for the previous pair, i.e.\ the multimodal posterior distribution of the inclination $\theta_\JN$ in the detector basis.
    In this case, though, both events are poorly localized, so the multimodality is not as pronounced as in the previous case, but it is still enough to cause a noticeable difference in the Gaussian shift.
\end{itemize}

Both of these outliers then represent a specific failure mode of Gaussian shifts due to bimodality in the parameter distribution.   
We conclude that the full parameter shift statistic is robust to this type of failure.

\begin{figure*}[!ht]
\centering
\includegraphics[width=\textwidth]{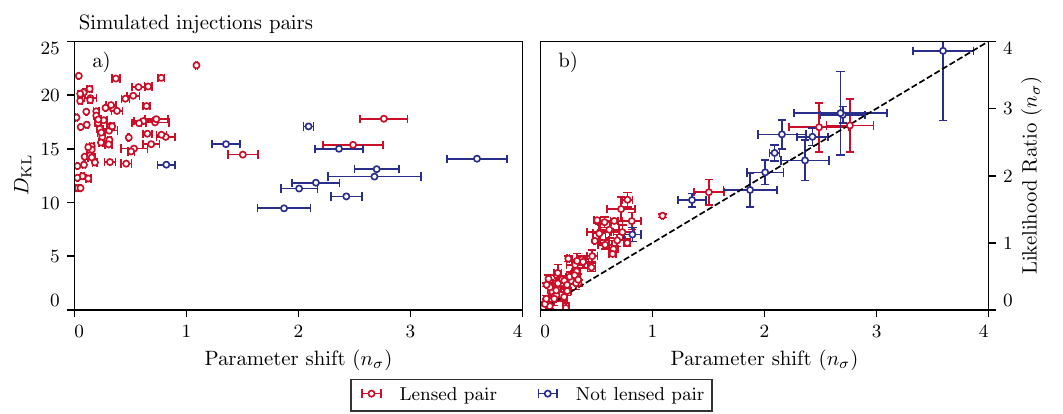}
\caption{\label{fig:sims.tensiometer_plane}
Lensing identification plane, panel (a), and likelihood ratio verification, panel (b), for simulated events pairs.
The KL divergence ($D_\KL$) quantifies the information content in the parameter difference distribution.  The parameter shift is the primary test of the lensing hypothesis but selected candidates should also pass the likelihood ratio test. 
Each point represents a different pair of simulated events. 
Different colors correspond to either lensed and not lensed pairs, as shown in the legend.
}
\end{figure*}

\subsection{Lensing candidates identification strategy}\label{sec:results.sims.lensing_plane}

Based on the examples in the GW injection catalog, we see that parameter shifts in the detector basis should be the primary indicator of consistency with lensing.  In the class of consistent events, the exact significance it assigns is less important since noisy events can be consistent with lensing because of their poorly constrained  parameters.  We therefore propose that the identification occur in the plane of parameter shift significance vs.\ KL information, $n_\sigma - D_\KL$.  
As parameter information increases, localization of the posterior in parameter space increases, making it more unlikely that two unrelated posteriors would show strong consistency of their parameter determinations due to random chance.
The likelihood ratio can be used as a final check that lensing is also a good fit to the data rather than only occupying a large fraction of the parameter posterior volume.  Gaussian shift can further flag specific types of non-Gaussianity when the primary indicators disagree and is automatically performed in the preselection process in any case. On the other hand, lensing analysis in the overlap basis can be omitted entirely.

In Fig.~\ref{fig:sims.tensiometer_plane} (a), we show the primary identification plane of parameter shift $n_\sigma - D_\KL$. 
Notice that the majority of the lensed pairs populate the $n_\sigma \lesssim 1$ and $D_\KL>15$ regime.   
The not-lensed pairs mostly have high $n_\sigma\gtrsim 2$, with many of the highest $D_\KL$ pairs falling outside the range of the plot, and those that have lower values tend to also have smaller $D_\KL$.    
Furthermore in Fig.\ \ref{fig:sims.tensiometer_plane} (b), we show that the likelihood ratio confirms that the majority of the lensed pairs are also good fits to lensing as we have already seen in Fig.~\ref{fig:tensiometer_like_ratio_smooth_vs_param_shift}. 

Fig.~\ref{fig:sims.tensiometer_plane} (a) suggests that the decision boundary for non-ambiguous lensing candidates should be diagonal in the $n_\sigma - D_\KL$ plane. Below this diagonal in the low $n_\sigma$, low $D_\KL$ corner while there could be lensed pairs there is simply not enough information to distinguish this small population from unlensed pairs, which dominate the total number.
On the other hand in the low $n_\sigma$, high $D_\KL$ corner we would have a region where the lensed pairs are well localized and separated from the not-lensed pairs, so they can be identified as promising lensing candidates.
In the intermediate region along the diagonal at moderate $n_\sigma\sim 2-3$, we have a region where both the lensed and unlensed pairs are not so well localized in parameter space, so they overlap in the $n_\sigma - D_\KL$ plane, as it is relatively more likely to encounter a noise fluctuation that brings the parameter determinations into or out of agreement.

Further injection population studies would be required to better quantify the lensing decision boundary in the $n_\sigma-D_\KL$ plane but this rough grouping into categories of promising lensing pairs, ambiguous pairs and unlikely lensing pairs is already useful for followup purposes which address the astrophysical considerations for lensing significance as we shall see next for the real GWTC-3 events.

\section{Lensing Candidates Identification in GWTC-3}\label{sec:results.real}

We now apply the lensing identification techniques developed on the NV and GW injections to the real GWTC-3 catalog where the lensing status is unknown a priori. 
We can however compare with past analyses and use this as a validation of our new methodologies.

\begin{figure*}[ht]
\centering
\includegraphics[width=\textwidth]{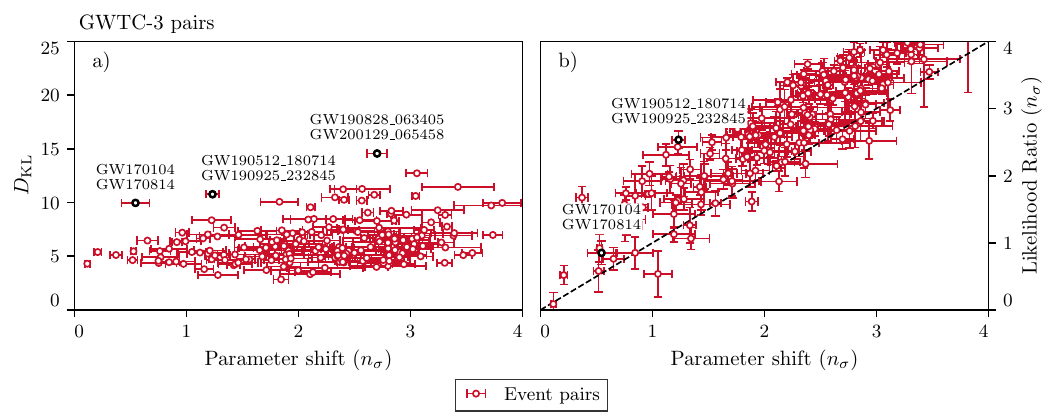}
\caption{\label{fig:real.lensing.plane}
Lensing identification plane, panel (a), and likelihood ratio verification, panel (b), for the GWTC-3 catalog.
Low parameter shift significance and high information indicate two possible lensing pair candidates and one ambiguous pair but the likelihood ratio test only passes for {\bf GW170104-GW170814}.  
}
\end{figure*}

\begin{figure}[!ht]
\centering
\includegraphics[width=\columnwidth]{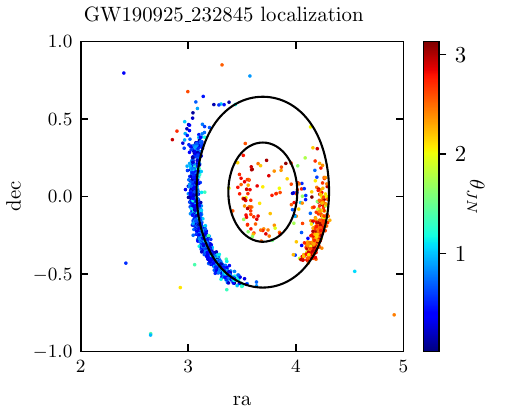}
\caption{\label{fig:GW190925.232845.localization}
Localization (ra, dec) posterior distributions for the GWTC-3 event \textbf{GW190925\textunderscore 232845} colored by inclination posterior values. The two distinct rings correspond to two different time delays in $\tau_{\rm HV}$.  Additionally the bimodal inclination parameter $\theta_\JN$ selects opposite sides of the two timing rings for each mode.
}
\end{figure}

\begin{figure}[!ht]
\centering
\includegraphics[width=\columnwidth]{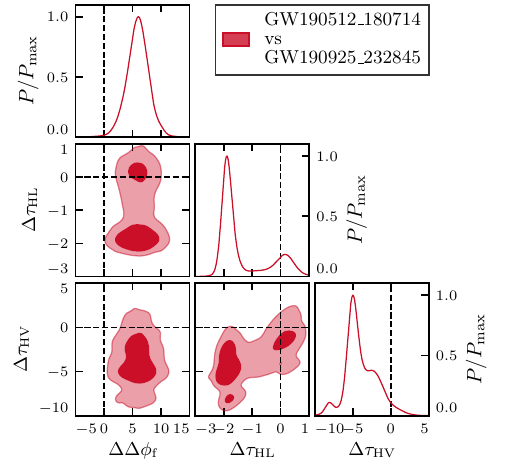}
\caption{\label{fig:GW190925.232845.triangle}
GWTC-3 lensing pair candidate \textbf{GW190512\textunderscore 180714-GW190925\textunderscore 232845}.
The multimodal localization rings of the latter in Fig.~\ref{fig:GW190925.232845.localization} lead to a bimodal time delay difference parameter $\Delta\tau_{\rm HL}$ in the detector basis, with one mode consistent with lensing (black dashed lines at zero parameter difference). Incompatibility in $\Delta\phi_f$ and the likelihood ratio it induces eliminates this pair as a promising lens candidate. 
}
\end{figure}

We begin by examining the  $n_\sigma-D_\KL$ parameter shift vs information plane, in Fig.~\ref{fig:real.lensing.plane} (a).
As we can see the real catalog is, as expected, lower overall SNR than the injection catalogs. 
As such we have several events at very low $n_\sigma$ but also low $D_\KL$. These are most likely not-lensed pairs that are simply poorly constrained by the data.
No pair lies in the convincingly high $D_\KL \gtrsim 15$, low $n_\sigma$ region, but two pairs, highlighted in Fig.~\ref{fig:real.lensing.plane} (a), stand out in the identification plane as the most promising candidates for followup:
\begin{itemize}
    \item {\bf GW170104-GW170814} this pair is known and was previously flagged in literature as a promising lensing candidate~\cite{Hannuksela:2019kle}.
    The parameter shift result is compatible with visual inspection of the posteriors, which show that all parameters broadly agree.
    In this case we also observe that the low lensing rejection significance is reinforced by the likelihood ratio in Fig.~\ref{fig:real.lensing.plane} (b) that gives a result that is compatible with parameter shifts.
    \item {\bf GW190512\textunderscore 180714-GW190925\textunderscore 232845} is a high information pair but the latter event is only seen by LIGO Hanford and Virgo, so its localization is not precise and allows two timing rings determined by $\tau_{\rm HV}$ (see Fig.~\ref{fig:GW190925.232845.localization}).
    This multi-modality reflects on the parameter difference distribution, that we show in Fig.~\ref{fig:GW190925.232845.triangle}, for relevant parameters in the detector basis of {\bf GW190512\textunderscore 180714}. The localization differences contribute little to the parameter shift or the likelihood ratio.
    However $\Delta\phi_f$ is individually discrepant more than the 2$\sigma$ level, despite the smaller overall significance of the total parameter shift. 
    The likelihood ratio excludes lensing at a significance of 2.5$\sigma$, mostly from the chirp mass result, since its distribution is nearly Gaussian.  
    This pair is therefore not a particularly promising strong lensing candidate because the well measured $\Delta\phi_f$ does not match and provides a poor likelihood ratio.
    Note that the likelihood ratio result, for this pair, agrees with the Gaussian shift result, that conservatively Gaussianizes the inclination multimodality while capturing the distance contribution due to $\Delta\phi_f$.
\end{itemize}

Finally, pairs with high KL and moderately high $n_\sigma\sim 2-3$ are also worth following up since the high SNR of these pairs may make small parameter shifts in an absolute sense appear significant. In this class, we have one event pair:

\begin{itemize}
    \item {\bf GW190828\textunderscore 063405-GW200129\textunderscore 065458} has high information content and a parameter shift that is only slightly above 2.5$\sigma$. This still makes it a possible lensing candidate, since, as we have seen in the injection catalog, lensed pairs that appear discrepant due to noise fluctuations can happen.
    However, the likelihood ratio rejection is very high, at 4.5$\sigma$ and mostly driven by multi-modal localization parameters, which indicates that the lensing hypothesis is not a good fit to the data, so we do not consider it a strong candidate.
\end{itemize}

We conclude that only GW170104-GW170814 is promising under our criteria.
This pair has been previously identified as interesting but further studies~\cite{Hannuksela:2019kle,Dai:2020tpj,Liu:2020par,LIGOScientific:2021izm,LIGOScientific:2023bwz,Ezquiaga:2023xfe} did not confirm the lensing hypothesis for this pair, based especially on selection effects and population modeling.

In App.~\ref{app:real.lensing.metrics} we further discuss event pairs that exhibit differences between methods and we comment on the types of non-Gaussianities that are responsible for these differences.

\section{Conclusions}\label{sec:conclusions}

In this paper we have developed, validated and applied a series of new machine-learning based methods to identify promising strong lensing candidates in gravitational wave (GW) catalogs. 
Those are characterized as repeated copies of an original merger signal that arrive at the detectors at different times and with different amplitudes and phases.

Our methods leverage the use of state-of-the-art normalizing flows (NF) models to perform several statistical calculations that are otherwise unfeasible with standard techniques. 
In particular we have shown: \emph{(i)} how to calculate the lensing information content of a pair of events; \emph{(ii)} how to quantify lensing consistency in parameter space by either calculating the significance of a parameter shift and likelihood ratio. 
Our methods differ from  previous approaches~\cite{Haris:2018vmn,Barsode:2024zwv} in that they do not require large simulations of (un)lensed events to calibrate the consistency with the lensing hypothesis, and extend those that did not need them~\cite{Ezquiaga:2023xfe} to include non-Gaussian statistical metrics.

To this end, we have developed NF models specifically tailored for GW posterior analysis. These models efficiently handle periodic parameters, which are common in GW parameter spaces (e.g.\ the reference phase or sky localization), and employ spline-based architectures to accurately capture multimodal and highly non-Gaussian distributions typical of GW events (e.g.\ due to the degeneracies between parameters such as the component masses or the distance and inclination).
We have found the performances of these models to be close to the ideal case with stringent Kolmogorov-Smirnov (KS) tests.
We publicly release the NF code as part of the \texttt{tensiometer} package~\cite{website:tensiometer}.

We have tested our techniques on three catalogs: one catalog consisting of different noise realizations of the same event to test their statistical properties; one catalog with both lensed and not-lensed simulated events to test  completeness and purity; and the third Gravitational Wave Transient Catalog (GWTC3) from the LIGO-Virgo-KAGRA Collaborations containing 86 binary black holes~\cite{LIGOScientific:2021djp}. 
Across all catalogs we have found that the detector parameter basis, introduced in~\cite{Ezquiaga:2023xfe}, has both more information and better identification than the parameter overlap basis~\cite{Hannuksela:2019kle}, because of the inclusion of Morse phase and the reduction of parameter degeneracies, especially relating to chirp mass.  
While Morse phase can~\cite{Barsode:2024zwv} be approximated by the addition of a reference phase, the detector basis achieves high information with fewer parameters.

We have found that information content plays an important role in enhancing the purity of the lensing candidate catalog.  
Events pairs that are highly localized in parameter space are less likely to be consistent by chance.
Retaining more information in the detector basis allows us to better identify pairs that are consistent with lensing, while rejecting pairs that are not lensed.

Across all catalogs we then proceed to define a lensing identification strategy that combines several insights. 
We perform a conservative preselection using Gaussian parameter shifts to reduce the computational cost without loss of lensing catalog completeness. This procedure approximates the distribution of GW parameter differences with a Gaussian distribution, allowing us to quickly identify pairs that are clearly inconsistent with lensing.
This allows us to trim down the number of pairs that we need to analyze with more expensive algorithms.

We then proceed to the training of NF models to calculate the significance of non-Gaussian parameter shifts. 
This method allows us to properly quantify the significance of a parameter difference, while taking into account the ubiquitous non-Gaussianity of GW posteriors. 
With simulated catalogs we have shown that this method assigns the correct statistical significance to events that are actually lensed. 
Events that are consistent with lensing are then identified by a low parameter shift significance and high information content.

We have discussed the extension of Wilks' theorem~\cite{Wilks:1938dza} to the case of the likelihood at maximum posterior to develop a likelihood ratio test statistic that quantifies the goodness of fit of the lensing hypothesis to the data.
This is then useful to reject poor fits to lensing when there are long but low probability density non-Gaussian tails that make parameter shifts seem likely.

When applied to the current GWTC-3 catalog, our method identifies the known lensing candidate, GW170104-GW170814, as the only significant one.
One other pair is identified by a small parameter shift, but is not confirmed by the likelihood ratio test statistic, which indicates that the lensing hypothesis is not a good fit to the data.

Further studies, with larger simulated catalogs, are needed to better characterize the lensing decision boundary in the plane of parameter shift significance vs information as well as the addition of astrophysical criteria and are left for future work.
Our methods could be extended to study other types of events such as binary neutron stars that in the future will be promising multi-messenger lensing sources~\cite{Smith:2025axx}.
They could also be extended to analyze sub-threshold GW lensing candidates that exhibit an even more complicated posterior distributions for their parameters.
Altogether, this works represents a methodological  step forward toward the goal of a robust statistical identification of the first lensed gravitational wave event.

\begin{acknowledgments}
We thank 
Andrea Amoretti,
Enzo Branchini,
Thomas Callister,
Daniel Holz,
Silvano Tosi,
Rico Lo,
Colm Talbot
for the helpful discussions and comments.
Computing resources were provided by the National Energy Research Scientific Computing Center (NERSC), a U.S. Department of Energy Office of Science User Facility operated under Contract No. DE-SC0009924, and by the University of Chicago Research Computing Center through the Kavli Institute for Cosmological Physics. 
G.C. and M.R. acknowledge financial support from the INFN InDark initiative. 
M.R. acknowledges financial support from the ICSC Spoke 2 under the grant NFL-GW.
W.H. was supported by U.S.\ Dept.\ of Energy contract DE-FG02-13ER41958 and the Simons Foundation. 
JME is supported by the European Union's Horizon 2020 research and innovation program under the Marie Sklodowska-Curie grant agreement No. 847523 INTERACTIONS, and by VILLUM FONDEN (grant no. 53101 and 37766). 
The Center of Gravity is a Center of Excellence funded by the Danish National Research Foundation under grant No. 184. 
The Tycho supercomputer hosted at the SCIENCE HPC center at the University of Copenhagen was used for supporting this work. 
This material is based upon work supported by NSF's LIGO Laboratory which is a major facility fully funded by the National Science Foundation.
\end{acknowledgments}

\appendix


\section{Mathematical Details}\label{App:Math.Details}

\subsection{Proofs of Sec~\ref{sec:information.content}}\label{App:KL.Proofs}

In the following, we demonstrate the validity of the data processing inequality (DPI) for the Kullback-Leibler (KL) divergence, Eq.~\eqref{Eq:DPI},  and we derive the formula for the KL divergence averaged over data realizations, Eq.~\eqref{Eq:avg.KL.volume}, both presented in Sec.~\ref{sec:information.content}.

Consider a non-invertible transformation $f:\ \theta \to \theta' = f(\theta)$.
We express the distributions $ P(\theta) $ and $ \Pi(\theta) $ in terms of the transformed variable $ \theta' $ and the conditional distributions $ P(\theta \mid \theta') $ and $ \Pi(\theta \mid \theta') $ as $P(\theta) = P(\theta')d\theta' P(\theta \mid \theta')$ and $\Pi(\theta) = \Pi(\theta')d\theta' \Pi(\theta \mid \theta')$.
Substituting into the KL divergence definition, Eq.\ \eqref{Eq:Def_KL_div}:
\begin{align}\label{Eq:kl_1}
{}&D_{\text{KL}}(P(\theta) \,||\, \Pi(\theta)) \\
={}& \int P(\theta) \log \left( \frac{P(\theta') P(\theta \mid \theta')}{\Pi(\theta') \Pi(\theta \mid \theta')} \right) \, d\theta \notag \\
={}&\int P(\theta) \log \left( \frac{P(\theta')}{\Pi(\theta')} \right) \, d\theta +\int P(\theta) \log \left( \frac{P(\theta \mid \theta')}{\Pi(\theta \mid \theta')} \right) \, d\theta. \notag
\end{align}
In going from $\theta \to \theta'$ through a not-one-to-one mapping we shall account for the fact that multiple 
$\theta$'s can map to the same $\theta'$. Thus the probability density over $\theta'$ results from summing over all possible $\theta$'s that could lead to a given $\theta'$. This is done by means of the conditional distribution:
 $P(\theta) d\theta \rightarrow P(\theta') \left(\int P(\theta | \theta') d\theta\right) \, d\theta'$.
Therefore we can simplify the first term in Eq.~\eqref{Eq:kl_1} as:
\begin{align}\label{Eq:kl_term1}
{}&\int P(\theta) \log \left( \frac{P(\theta')}{\Pi(\theta')} \right) \, d\theta = \notag \\
{}&= \int \left( \int P(\theta \mid \theta') \, d\theta \right) P(\theta') \log \left( \frac{P(\theta')}{\Pi(\theta')} \right) \, d\theta'\notag \\
{}&\equiv D_{\text{KL}}(P(\theta') \,||\, \Pi(\theta')),
\end{align}
where we have identified the KL divergence between the two distributions in the transformed parameters basis in the last step, being $ P(\theta \mid \theta')$ a normalized conditional distribution.
We analogously perform a change of variable in the second term of Eq.~\eqref{Eq:kl_1}:
\begin{align}\label{Eq:kl_term2}
{}&\int P(\theta) \log \left( \frac{P(\theta \mid \theta')}{\Pi(\theta \mid \theta')} \right) \, d\theta = \notag \\
{}& = \int P(\theta') \left( \int P(\theta \mid \theta') \log \left( \frac{P(\theta \mid \theta')}{\Pi(\theta \mid \theta')} \right) \, d\theta \right) d\theta' \notag \\
{}& \equiv \int P(\theta') D_{\text{KL}}(P(\theta \mid \theta') \,||\, \Pi(\theta \mid \theta')) \, d\theta'\notag \\
{}& \equiv \mathbb{E}_{\theta'} \left[ D_{\text{KL}}(P(\theta \mid \theta') \,\|\, \Pi(\theta \mid \theta')) \right],
\end{align}
where eventually we have introduced the expectation value of the conditional KL divergence over the probability distribution of $\theta'$, denoted by the operator $\mathbb{E}_{\theta'}[\cdot]$.
Combining Eq.~\eqref{Eq:kl_term1} and Eq.~\eqref{Eq:kl_term2} into Eq.~\eqref{Eq:kl_1} we find the expression for the KL-Decomposition Formula or KL-chain rule:
\begin{align}\label{Eq:chain_rule}
  D_{\text{KL}}(P(\theta) \,||\, \Pi(\theta)) ={}&  \\
 D_{\text{KL}}(P(\theta') \,||\, \Pi(\theta')) \\
&\!\!\! + \mathbb{E}_{\theta'} \left[ D_{\text{KL}}(P(\theta \mid \theta') \,\|\, \Pi(\theta \mid \theta')) \right].\notag
\end{align}
Since the KL divergence is always non-negative, also the expectation value for KL in the above expression must be
non-negative.
Thus, we have verified the DPI Eq.~\eqref{Eq:DPI}.

Now let us consider the special case in which we have a bijective transformation $g:\ \theta \to \tilde{\theta} = g(\theta)$. Since the determinant of the Jacobian of the transformation $g$ never vanishes, then we have $P(\theta)d\theta = P(\tilde{\theta}) d\tilde{\theta}$, and of course the same would happen for the prior distribution. In this case we would find:
\begin{align} \label{Eq:KL_div_inv_reparam}
D_{\text{KL}}(P(\tilde{\theta}) \,||\, \Pi(\tilde{\theta})) = D_{\text{KL}}(P(\theta) \,||\, \Pi(\theta)).
\end{align}
This invariance under bijective transformations property implies that the KL divergence is a coordinate-free measure of the difference between two probability distributions.

We now verify the linear relationship between the average KL
divergence over data realizations and the posterior/prior volume ratio in the context
of the Gaussian Linear Model, Eq.~\eqref{Eq:avg.KL.volume}.
We start by taking the average of Eq.~(\ref{Eq:KLGaussians_volume}) over  realizations of the data vector $x$: 
\begin{align} \label{Eq:KLGaussians_average}
2\langle D_{\rm KL}(P \,||\, \Pi)\rangle_x ={}& \langle (\theta_{\Pi} - \theta_p)^T \mathcal{C}_{\Pi}^{-1} (\theta_{\Pi} - \theta_p)\rangle_x \nonumber\\ 
&+ 2 \log ( V_p^{-1}V_{\Pi})  - N_{\rm eff}.
\end{align}
We know that 
\begin{align} \label{Eq:expectation_of_quadratic_form}
\mathbb{E}_x[X^T A X] \equiv \mathbb{E}_x[X]^T A \mathbb{E}_x[X] + {\rm tr}[A\text{Var}_x(X)]
\end{align}
so we can apply this formula to the first term in \eqref{Eq:KLGaussians_average} where we have $X=\theta_{\Pi}(x) - \theta_p (x)$, $A=\mathcal{C}_\Pi^{-1}$ and $\text{Var}_x(x) = \text{Var}_x(\theta_{\Pi} - \theta_p)$.
We can make the dependence on the data explicit by making use of the GLM formalism ~\cite{Raveri:2018wln}:
\begin{align} \label{Eq:GLMParameterDifferences}
\theta_p - \theta_\Pi = \mathcal{C}_p M^T \Sigma^{-1}(x - m_\Pi)
\end{align}
where we have assumed Gaussian distributed data with likelihood $\mathcal{L}_x(\theta) \equiv \mathcal{N}_d(x; m(\theta), \Sigma)$, $m_\Pi \equiv m(\theta_\Pi)$ is the expansion point and $M \equiv \partial m / \partial \theta |_{\theta_\Pi}$ is the linear model Jacobian, evaluated at the expansion point.
Since $\mathbb{E}_x[\theta_{\Pi} - \theta_p]$ is proportional to $\mathbb{E}_x[x - m_{\Pi}]$ and $\mathbb{E}_x[x]= m_{\Pi}$, the term corresponding to the first part in \eqref{Eq:expectation_of_quadratic_form} vanishes, therefore we obtain:
\begin{align} \label{Eq:KLGaussians_average_dataspace}
2\langle D_{\rm KL}(P \,||\, \Pi)\rangle_x={}& {\rm tr}[ \mathcal{C}_\Pi^{-1} \text{Var}_x(\theta_{\Pi} - \theta_p)]\notag\\ 
&+2 \log ( V_p^{-1}V_{\Pi})  - N_{\rm eff}.
\end{align}
Employing the formula for the variance of a linear form $\text{Var}_x[AX]=A\text{Var}_x [X]A^T$, we can write
\begin{align} \label{Eq:variance1}
\text{Var}_x[\theta_{\Pi} - \theta_p]&=\text{Var}_x[\mathcal{C}_p M^T\Sigma^{-1}(x-m_{\Pi})] \\
&=\mathcal{C}_p M^T\Sigma^{-1}\text{Var}_x[x-m_{\Pi}](\mathcal{C}_p M^T\Sigma^{-1})^T. \notag
\end{align}
We know that in the GLM the evidence is Gaussian and given by:
\begin{align} \label{Eq:GLMEvidenceDistribution}
\mathcal{E}(x) = \mathcal{N}_d(x; m_\Pi, \Sigma + M\mathcal{C}_\Pi M^T) 
\end{align}
therefore we identify $\text{Var}_x[x-m_{\Pi}]$ as the variance of the evidence in the GLM: $\text{Var}_x[x-m_{\Pi}]=\Sigma + M\mathcal{C}_\Pi M^T$. Now, recalling that in the linear model, data and parameters are related by a non-invertible linear transformation, so that the projection of data covariance over parameter space is given by $\mathcal{C} = (M^T \Sigma^{-1} M)^{-1}$, we can rewrite Eq.~\eqref{Eq:variance1}:
\begin{align} \label{Eq:variance2}
\text{Var}_x[\theta_{\Pi} - \theta_p]&=\mathcal{C}_p \mathcal{C}^{-1}(\mathbb{I}_{N\times N} + \mathcal{C}_{\Pi}\mathcal{C}^{-1})\mathcal{C}_p^T.
\end{align}
We know that the relation $\mathcal{C}_p^{-1} = \mathcal{C}^{-1} + \mathcal{C}_\Pi^{-1}$ holds between covariance matrices, therefore we obtain
\begin{align} \label{Eq:variance3}
\text{Var}_x[\theta_{\Pi} - \theta_p]=\mathcal{C}_{\Pi}-\mathcal{C}_p.
\end{align}
Inserting \eqref{Eq:variance3} in \eqref{Eq:KLGaussians_average_dataspace} and employing the definition of $N_{\rm eff}$ \eqref{Eq:Neff}, we conclude
\begin{align}
\langle D_{\rm KL}(P \,||\, \Pi)\rangle_x= \log V_{\Pi} - \log V_p 
\end{align}
which demonstrates Eq.~\eqref{Eq:avg.KL.volume} and describes the relation between parameter volumes and the KL divergence of posterior after prior distributions.  Note that the posterior distribution always occupies a smaller volume than the prior, ensuring that the quantity $\langle D_{\rm KL}(P | \Pi)\rangle_x$ remains positive.
Additionally, the right-hand side is invariant under linear transformations. Because in the GLM framework all transformations are linear to maintain Gaussianity, this property reflects the broader parameter invariance characteristic of the KL divergence itself.

\subsection{Proofs of Sec~\ref{sec:likelihood.ratio}} \label{App:likelihood.ratio.Proofs}
In this appendix, we provide additional details concerning the likelihood ratio test statistic introduced in Sec.~\ref{sec:likelihood.ratio}, along with the method used to approximate its distribution.

Consider the likelihood ratio test statistic defined in Eq.~\eqref{Eq:Like_ratio_test_def}. In the GLM formalism it can be rewritten as:
\begin{align} \label{Eq:Like_ratio_test_GLM}
Q_\mathcal{L} \equiv{}& (\Delta \theta_{\rm f} - \Delta \theta_\ML)^T\mathcal{C}^{-1}(\Delta \theta_{\rm f} - \Delta \theta_\ML)\\ 
&-(\Delta \theta_{\rm MAP} - \Delta \theta_\ML)^T\mathcal{C}^{-1}(\Delta \theta_{\rm MAP} - \Delta \theta_\ML)\notag
\end{align}
where $\Delta \theta_{\rm f}$ is a fixed parameter difference and $\Delta \theta_{\rm MAP}$
is the parameter difference at maximum posterior, which within the GLM, corresponds to the posterior mean, while $\Delta \theta_\ML$ and $\mathcal{C}$ are maximum likelihood parameters difference and parameter difference covariance in the GLM framework~\cite{Raveri:2018wln}.
This quantity can be expressed as a block quadratic form $Q_\mathcal{L}\equiv X^T A X$ with the data vector $X$
\begin{align}
X= \begin{pmatrix} \Delta \theta_{\rm f} - \Delta \theta_\ML \\ \Delta \theta_{\rm MAP} - \Delta \theta_\ML \end{pmatrix},
\end{align}
which is Gaussian distributed in data space, and the block diagonal matrix $A$
\begin{align}
A=\begin{pmatrix}  \mathcal{C}^{-1} & 0 \\ 0 & -\mathcal{C}^{-1}
\end{pmatrix}.
\end{align}
The first two moments of a generic quadratic form $ Q\equiv X^T A X $ for $X\sim \mathcal{N}_d(x; \mu, \Sigma)$ are:
\begin{align} \label{EqApp:QuadraticFormMomenta}
\langle Q\rangle_x =\,& {\rm tr} [A\Sigma]+\mu^T A \mu \,, \nonumber \\
\text{Var}_x[Q] =\,& 2\,{\rm tr} [(A\Sigma)^2] +4\mu^TA \Sigma A\mu \,.
\end{align}
To compute mean and covariance of $X$, 
note that in the GLM formalism $\mathbb{E}_x[\Delta \theta_{\rm ML}]=\mathbb{E}_x[\Delta \theta_{\rm MAP}] = \Delta \theta_{\Pi}$, leading to:
\begin{align} \label{Eq:X_moments}
\mu=\begin{pmatrix} \Delta \theta_{\rm f} - \Delta \theta_{\Pi} \\ 0 \end{pmatrix} \\
\Sigma=\begin{pmatrix} \mathcal{C} & \mathcal{C} - \mathcal{C}_p\\
\mathcal{C} - \mathcal{C}_p & \mathcal{C} - \mathcal{C}_p\end{pmatrix}.
\end{align}
In our analysis the likelihood ratio is evaluated with respect to $\Delta \theta_{\rm f}=0$.  Furthermore, Wilks' theorem ensures that under the null hypothesis, the expected value of the fitted parameter difference matches that of the posterior mean: $\mathbb{E}_x[\Delta \theta_{\rm f}] = \mathbb{E}_x[\Delta \theta_{\rm MAP}]$, i.e. $\Delta \theta_{\rm f} = \Delta \theta_{\Pi}$. This choice yields a null mean $\mu = 0$, allowing us to treat $Q_\mathcal{L}$ as a central quadratic form. 
Hence the mean and the variance of the quadratic form can be simplified and we find moments for the distribution of our test statistic $Q_\mathcal{L}$ are:
\begin{align} \label{Eq:like_ratio_moments}
\langle Q_\mathcal{L} \rangle_x={}&  {\rm tr}\left[A\Sigma\right]=N_{\rm eff}\,,\\
\text{Var}_x[Q_\mathcal{L}]={}&  2\,{\rm tr}\left[(A\Sigma)^2\right] \notag\\
={}& 2N-2\,{\rm tr}\left[\left(\mathcal{C}_{\Pi}^{-1}\mathcal{C}_p\right)^2\right]\,.
\end{align}
Note that here $N_{\rm eff}$ is the effective number of degrees of freedom of the posterior parameter difference distribution $P(\Delta \theta \equiv \theta_1 - \theta_2)$, as well as all covariances refer to parameter difference posterior and prior distributions. If the two GW events posteriors are approximated by two uncorrelated Gaussian distributions with mean $\theta_{p_1}$ and $\theta_{p_2}$ and covariance $\mathcal{C}_{p_1}$ and $\mathcal{C}_{p_2}$ respectively,
then the distribution of parameter differences is Gaussian with mean $\langle \Delta \theta \rangle =\theta_{p_1}-\theta_{p_2}$ and covariance $\mathcal{C}_{p_1} + \mathcal{C}_{p_2}$. 
 In this case we have, following the definition \eqref{Eq:Neff}
\begin{align} \label{Eq:Neff_param_diff}
N_{\rm eff} (\Delta \theta)
&= N -{\rm tr}[ \mathcal{C}_\Pi^{-1}(\Delta \theta)\mathcal{C}_p(\Delta \theta) ] \notag\\
&=N-\frac{1}{2}{\rm tr}\left[C_\Pi^{-1}\left(\mathcal{C}_{p_1} + \mathcal{C}_{p_2}\right)\right]\notag\\
&=N-\frac{1}{2}{\rm tr}\left[C_\Pi^{-1}\mathcal{C}_{p_1}\right]-\frac{1}{2}{\rm tr}\left[C_\Pi^{-1}\mathcal{C}_{p_2}\right]\notag\\
&=\frac{1}{2}\left[N_{\rm eff}(\theta_1)+N_{\rm eff}(\theta_2)\right],
\end{align}
so we can conclude that the number of effective degrees of freedom of the difference distribution is the average of the two single ones.

\begin{figure*}[!ht]
\centering
\includegraphics[width=\textwidth]{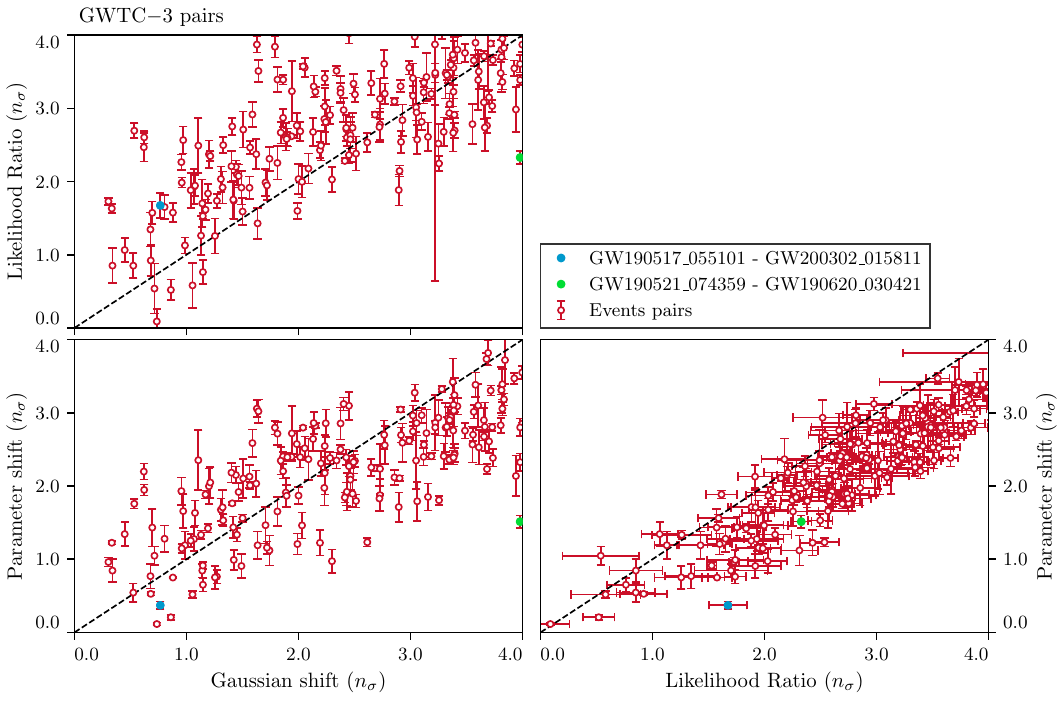}
\caption{\label{fig:real.lensing.metrics}
Comparison between lensing metrics for the GWTC-3 catalog: Gaussian and non-Gaussian parameter shifts significance, Likelihood ratio significance. 
Each point represents a different pair of events. 
Error bars correspond to flow model variance whereas for the Gaussian shift no error estimates are available. 
Different colors correspond to either lensed and not lensed pairs, as shown in the legend.
}
\end{figure*}

We can decompose $Q_\mathcal{L}$ as:
\begin{align} \label{Eq:QuadraticForm_decomposition}
  Q_\mathcal{L}\equiv X^T A X = \sum_i \lambda_i U_i^2 
\end{align}
where we have defined the non-zero eigenvalues of $A\Sigma$ as $\lambda_i$ and decomposed the form so that the random variables $U_i \sim \mathcal{N}_1(u; 0, 1)$ are distributed as a unit Gaussian. Since all the eigenvalues of $A\Sigma$ are non negative, then analytic expressions for the probability density of $Q_\mathcal{L}$ exist - it generally corresponds to a weighted sum of chi-square variables with weights given by the eigenvalues of $A\Sigma$ - and probabilities can be computed with dedicated algorithms~\cite{QuadraticFormsinRandomVariables,liu2009computing}, once the eigenvalues are obtained. 
 
Alternately, the distribution of a quadratic form can  be approximated by that of a chi squared variable matching some of the moments of the quadratic form itself~\cite{Patnaik}.
The simplest and most commonly used version is Patnaik's First approximation, which approximates the distribution of $Q_\mathcal{L}$  by a chi-square distribution that matches the mean of the original quadratic form, as given in Eq.~\eqref{Eq:like_ratio_moments}:
\begin{align} \label{Eq:like_ratio_Patnaiks1}
 Q_\mathcal{L} & \sim  \chi^2(\langle Q_\mathcal{L}\rangle_x) \sim \chi^2(  {\rm tr}[A\Sigma] )\nonumber\\
& \hspace{0.1cm} \Rightarrow \hspace{0.1cm} Q_\mathcal{L} \sim \chi^2(N_{\rm eff} ) \,.
\end{align}
Notice that the number of degrees of freedom of the chi square distribution is usually not integer.

In the limit where parameters are fully constrained by data, i.e. $N_{\rm eff}\to N$,  $\mathcal{C}_{\Pi}^{-1}\mathcal{C}_p \to 0$ and we find $\text{Var}_x[Q_\mathcal{L}]\to 2N$, which corresponds to the variance of a $\chi^2$-distribution with $N$ degrees of freedom, confirming the approximation’s consistency in the strongly data-driven regime.  On the other hand, when parameters are fully prior dominated and no constraints come from data, i.e. $N_{\rm eff}=0$, $\mathcal{C}_p\to\mathcal{C}_{\Pi}$, so the variance approaches zero, as for a $\chi^2$-distribution with zero degrees of freedom. Here the approximation correctly reflects that the test statistic carries no information to distinguish posterior from prior.\\

An important property of this approximation is that the variance of the quadratic form $Q_\mathcal{L}$ is always
$\text{Var}_x[Q_\mathcal{L}]\geq2N_{\rm eff}$, since the eigenvalues of $\mathcal{C}_{\Pi}^{-1}\mathcal{C}_p$ lie between $0$ and $1$. This means that the variance of our test statistic is always greater than the variance of the approximating chi squared distribution with $N_{\rm eff}$ degrees of freedom. Consequently, we conclude that this kind of approximation for $Q_\mathcal{L}$, used in the main text, can in principle overestimate the significance of rejection of the lensing hypothesis.  However in practice,
this overestimate is typically negligible since the number of partially constrained parameters is usually small, and the approximation is exact in the case where parameters are fully constrained by the data or the prior.

\section{Lensing significance with different estimators for GWTC-3}\label{app:real.lensing.metrics}

As it was done with simulated injections in Sec.~\ref{sec:results.different_estimators}, it is useful to check when the three tension metrics disagree for events pairs from the real catalog. 
In Fig.~\ref{fig:real.lensing.metrics}, we compare the parameter shift, likelihood ratio and Gaussian shift estimators of lensing significance.  While the estimators still show a high degree of correlation, the scatter between them is notably larger than in the injection catalogs.  This is mainly due to the lower signal to noise and the many pairs with effectively only two detectors, which leads to a higher level of non-Gaussianity in the parameter posteriors.  Another notable difference is that Gaussian shifts tends to report lower significance than likelihood ratios.  This can partially be attributed to multimodal distributions inflating the covariance and hence the error metric used by the Gaussian shift. 

Once again, it is useful to examine the outlier cases where the techniques differ the most:

    \begin{figure}[!ht]
    \centering
    \includegraphics[width=\columnwidth]{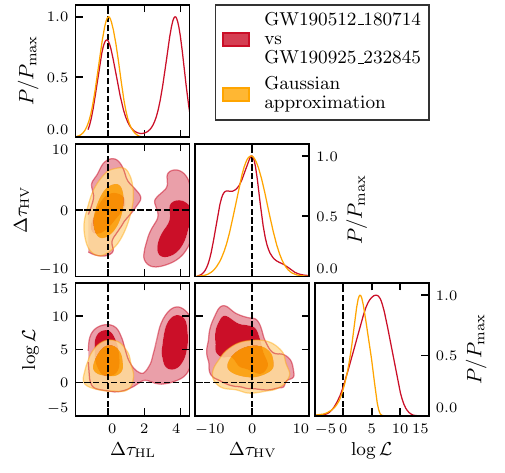}
    \caption{\label{fig:real.outlier.1}
    Posterior distributions for parameter differences of {\bf GW190517\textunderscore 055101-GW200302\textunderscore 015811}.  The multimodal localization parameters shifts $\Delta\tau_{\rm HL},\Delta\tau_{\rm HV}$ reveal one mode, including its Gaussianized approximation, thiat is consistent with lensing but the other mode encompasses the MAP and has higher likelihood.  This causes the likelihood ratio to have a higher $n_\sigma$ than parameter shifts.
    }
    \end{figure}
    \begin{figure}[!ht]
    \centering
    \includegraphics[width=\columnwidth]{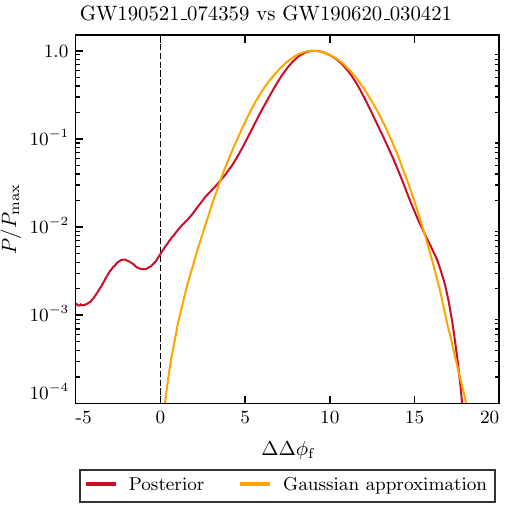}
    \caption{\label{fig:real.outlier.2}
    Posterior distribution for the shift in $\Delta\phi_f$
     of {\bf GW190521\textunderscore 074359-GW190620\textunderscore 030421}.  The Gaussian approximation underestimates the probability of zero shift leading to a larger $n_\sigma$ than reported by the likelihood ratio.   The parameter shift $n_\sigma$ is even smaller due to the long low probability tail to smaller $\Delta\Delta\phi_f$ that encompasses a large parameter volume. 
    }
    \end{figure}   
    
\begin{itemize}
    \item {\bf GW190517\textunderscore 055101-GW200302\textunderscore 015811}: 
    this outlier is interesting because it shows a significant difference between the likelihood ratio and parameter shift estimators, both Gaussian and non-Gaussian.
    This pair is one of the most consistent in the low-KL region and, for this reason, shows pronounced non-Gaussianities in most parameters and multimodality in localization.

    In Fig.~\ref{fig:real.outlier.1} we show localization and the distribution of likelihood values.
    We see that the distribution has one peak that sits almost exactly on zero parameter shift and another peak that is further away from zero. Although Gaussianization would normally merge distinct modes, the {\tt phazap} algorithm conservatively separates the localization modes above vs.\ below the HLV detector plane and Gaussianizes the one that is most consistent with lensing.
    The presence of a peak very close to zero shift then reports a small $n_\sigma$ value for both types of parameter shifts.
    The increased likelihood ratio is due to the fact that the maximum posterior is in fact in the second mode away from zero and has a larger likelihood as well.
    \item {\bf GW190521\textunderscore 074359-GW190620\textunderscore 030421}: 
    this outlier is a significant incompatibility with lensing for Gaussian shift while the parameter shift is consistent with lensing and the likelihood ratio falls in between the two. 
    In Fig.~\ref{fig:real.outlier.2} we show the distribution of differences in $\Delta \phi_f$.
    The most notable feature is the slow decay of the tail of the distribution. The Gaussian approximation decays very quickly while the real distribution falls off and reaches zero at a significantly higher value and moreover has a long tail in parameter space to even lower values.  Thus the Gaussian approximation overestimates the poor fit that the likelihood ratio reports as 2.4$\sigma$ which itself is larger than the parameter shift significance of 1.5$\sigma$ due to the large parameter volume of the tail.
    Other parameters are broadly in agreement and this raises the compatibility of the two distributions to the level reported by the other two estimators.
\end{itemize}

These outliers point towards the same conclusions as the main text. The estimators disagree because of  non-Gaussianities in the parameter difference distribution and that though parameter shifts should be the primary indicator, the likelihood ratio provides a secondary test that can further eliminate unlikely lens pairs.

\vfill
\bibliographystyle{apsrev4-1}
\bibliography{biblio}
\end{document}